\documentclass[10pt]{article} % For LaTeX2e
\usepackage[preprint]{tmlr}
\usepackage{hyperref}
\usepackage{url}
\usepackage[utf8]{inputenc}
\usepackage{amsfonts}
\usepackage{mathtools}
\usepackage{graphicx}
\usepackage{enumitem}   
\usepackage{amssymb,amsmath,amsthm}
\usepackage{nicematrix}
\usepackage{multirow}
\usepackage{amssymb}
\usepackage{subcaption}
\usepackage{accents}
\usepackage{xspace}
\usepackage{dsfont}

\newcommand{\review}[1]{{\color{black} #1}}
\usepackage{ulem}
\usepackage{cleveref}

\crefname{equation}{}{}

\newcommand{\signalset}{\mathcal{X}}
\DeclareMathOperator*{\signc}{sign}
\DeclareMathOperator*{\diam}{diam}
\newcommand{\sign}[1]{\signc\left(#1\right)}

\newcommand{\ntransf}{G}
\newcommand{\Prob}{\mathbb{P}}

\newcommand{\red}[1]{\textcolor{black}{#1}}
\newcommand{\corr}[1]{\textcolor{black}{#1}}
\newcommand{\R}[1]{\mathbb{R}^{#1}}

\newcommand{\ball}[1]{B_{\epsilon}{(#1)}}
\newcommand{\Sp}[1]{\mathbb{S}^{#1}}

\newcommand{\concatA}{\bar{A}}
\newcommand{\proposed}{SSBM}

\newcommand{\rk}[1]{\text{rank}(#1)}
\newcommand{\bdim}[1]{\operatorname{boxdim}\left(#1\right)}

\DeclareMathOperator*{\argmin}{arg\,min}
\DeclareMathOperator*{\boxdim}{boxdim}

\newtheorem{theorem}{Theorem}
\newtheorem{corollary}[theorem]{Corollary}
\newtheorem{lemma}[theorem]{Lemma}
\newtheorem{conjecture}[theorem]{Conjecture}
\newtheorem{proposition}[theorem]{Proposition}

\theoremstyle{definition}
\newtheorem{example}{Example}[section]
\newtheorem{definition}{Definition}[section]
\newtheorem{assumption}{Assumption}
\newcommand{\bb}{\mathbb}
\newcommand{\cl}{\mathcal}
\newcommand{\ts}{\textstyle}

\newcommand{\ie}{\emph{i.e.}, }
\newcommand{\eg}{\emph{e.g.}, }
\newcommand{\iid}{%
  \ifmmode% math mode
  \mathrm{i.i.d.}%
  \else%
  i.i.d.\@\xspace%
  \fi%
}

\title{Learning to reconstruct signals\\ from binary measurements alone}

\author{\name Juli\'an Tachella \email julian.tachella@cnrs.fr \\
      \addr Physics Laboratory\\
      CNRS \& École Normale Supérieure de Lyon
      \AND
      \name Laurent Jacques \email laurent.jacques@uclouvain.be\\
      \addr ICTEAM \\
      UCLouvain}

\def\month{11}  % Insert correct month for camera-ready version
\def\year{2023} % Insert correct year for camera-ready version
\def\openreview{\url{https://openreview.net/forum?id=ioFIAQOBOS}} % Insert correct link to OpenReview for camera-ready version

\begin{document}
\maketitle

\begin{abstract}
    Recent advances in unsupervised learning have highlighted the possibility of learning to reconstruct signals from noisy and incomplete linear measurements alone. These methods play a key role in medical and scientific imaging and sensing, where ground truth data is often scarce or difficult to obtain. However, in practice measurements are not only noisy and incomplete but also quantized. 
    Here we explore the extreme case of learning from binary observations and provide necessary and sufficient conditions on the number of measurements required for identifying a set of signals from incomplete binary data. Our results are complementary to existing bounds on signal recovery from binary measurements. Furthermore, we introduce a novel self-supervised learning approach\red{, which we name \proposed,} that only requires binary data for training. \red{We demonstrate in a series of experiments with real datasets that \proposed~performs} on par with supervised learning and outperforms sparse reconstruction methods with a fixed wavelet basis by a large margin.
\end{abstract}

\section{Introduction}
%\LJ{General comments:
%\begin{itemize}
%    \item I'm wondering if it would not be worth introducing $\cl X_\delta$ in the Introduction in order to better explain what we mean by approximate signal set identification.
%    \item I guess we should still improve our explanations on how knowing/estimating a signal set is connected to "learning to reconstruct", \ie to estimate $f_\theta$. After a new reading of the whole paper, I still find this unclear, and I guess reviewers could think the same. I discuss a possible storytelling in the beginning of Sec. 4.
%    \item I've changed a bit the formulation of our contributions in the end of the Introduction to make them closer to what we really show (at least to my understanding)
%    \item I have renamed the matrix $M$ stacking all measurement matrices $A_1, ..., A_G$ into $\concatA$.
%    \item I have an idea to simplify a bit the notations: writing $B(x)$ for \emph{b}inary mapping $\sign{A x}$, $B_g(x)$ for $\sign{A_g x}$, and $\bar B(x)$ for the one of $\sign{\concatA(x)}$. I didn't implement it. Note that $\hat{\cl X} = \bigcap_g [B_g^{-1} \circ B_g (\cl X)]$.   
%    \end{itemize}}
Continuous signals have to be quantized in order to be represented digitally with a limited number of bits in a computer.
In many real-world applications, such as radar~\citep{alberti1991time}, wireless sensor networks~\citep{chen2015amplitude}, and recommender systems~\citep{davenport2014onebit}, the measured data is quantized with just a few bits per observation. The extreme case of quantization corresponds to observing a single bit per measurement.
For example, single-photon detectors record the presence or absence of photons at each measurement cycle~\citep{kirmani2014first}, and recommendation systems often observe a binary measurement of users' preferences only (\eg via thumbs up or down).

The binary sensing problem is formalized as follows: we observe binary measurements $y\in \{-1,1\}^{m}$ of a signal $x\in \signalset \subset \Sp{n-1}$ with unit norm\footnote{Note that the sensing model in~\Cref{eq: onebit model} provides no information about the norm of $x$, so it is commonly assumed that signals verify $\|x\|=1$. } via the following forward model
\begin{equation} \label{eq: onebit model}
    y = \sign{A x}
\end{equation}
where $A\in \R{m\times n}$ is a linear forward operator.
Recovering the signal from the measurements is an ill-posed inverse problem since there are many signals $x\in \Sp{n-1}$ that are consistent with a given measurement vector $y$. Moreover, often the measurement matrix is incomplete $m<n$, \eg as in one-bit compressed sensing~\citep{jacques2013robust}, which makes the signal recovery problem even more challenging.

It is possible to obtain a good estimation of $x$ despite the binary quantization, if the set of plausible signals $\signalset$ is low-dimensional~\citep{bourrier2014fundamental}, \ie if it occupies a small portion of the ambient space $\Sp{n-1}$. A popular approach is to assume that $\signalset$ is a single linear subspace or a union of subspaces~\citep{jacques2013robust}, imposing sparsity over a known dictionary. For example, the well-known total variation regularization assumes that the gradients of the signal are sparse~\citep{rudin1992nonlinear}. However, in real-world settings, the set of signals $\signalset$ is generally unknown, and sparsity assumptions on an arbitrary dictionary yield a loose description of the true set $\signalset$, negatively impacting the quality of reconstructions obtained under this assumption. This limitation can be overcome by learning the reconstruction mapping $y\mapsto x$ (\eg with a deep neural network) directly from $N$ pairs of measurements and associated signals---\ie a supervised learning scenario with a labeled dataset $\{(y_i,x_i)\}_{i=1}^N$ with $N$ assumed sufficiently large. While this learning-based approach generally obtains state-of-the-art performance, it is often impractical since it can be very expensive or even impossible to obtain ground-truth signals $x_i$ for training. For example, recommender systems generally do not have access to high-resolution user ratings on all items for training.

 \begin{figure}[t]
\centering
\includegraphics[width=1\textwidth]{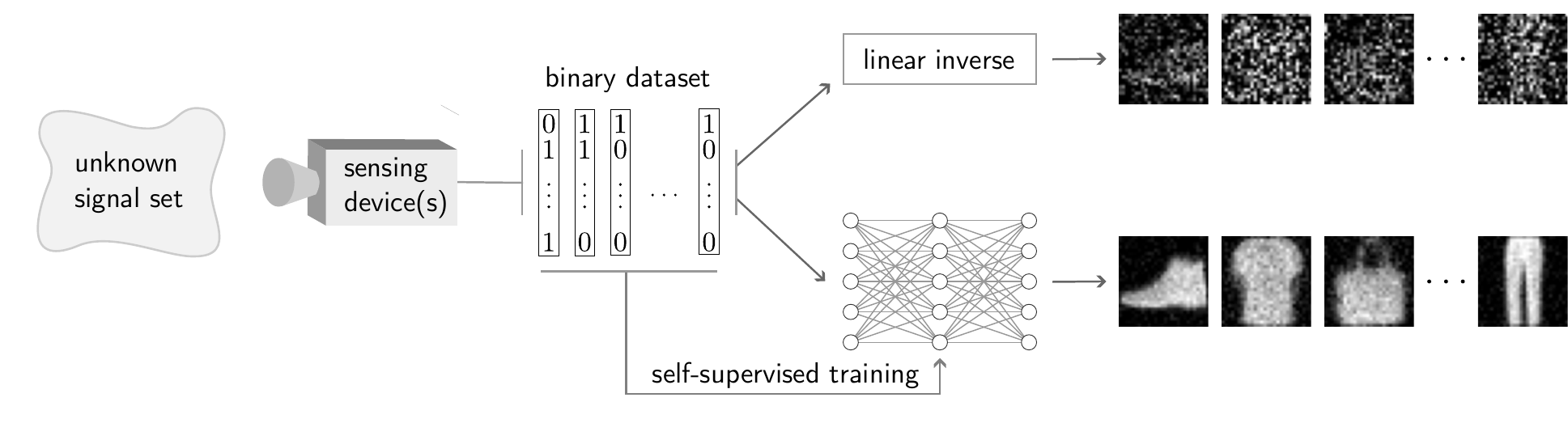}
\caption{ We propose a method for learning to reconstruct binary measurement observations, using only the binary observations themselves for training. The learned reconstruction function can discover unseen patterns in the data (in this case the clothes of fashionMNIST - see the experiments in~\Cref{sec: experiments}),  
 \red{which cannot be recognized in the standard linear reconstructions (no learning). We also provide theoretical bounds that characterize how well we can expect to learn the set of signals from binary measurement data alone.}}
\label{fig:schematic} 
\end{figure}

In this paper, we investigate the \red{problems of identifying the signal set and learning reconstruction mapping} using a dataset of binary measurements only $\{y_i\}_{i=1}^N$. 
%In this setting, if the measurement process is (over)complete $m\geq n$ and no prior information is known about $\signalset$, learning is not required, as the optimal reconstruction mapping corresponds to standard consistency reconstruction algorithms, \eg as described in~\citep{goyal1998quantized}.
In this setting, if the measurement process is incomplete $m<n$, the matrix $A$ has a non-trivial nullspace and there is no information in the measurement data about the set of signals $\signalset$ in the nullspace~\citep{chen2021equivariant}. \red{As a consequence, there is not enough information for learning the reconstruction function either. For example, the trivial pseudo-inverse reconstruction $f(y)  = A^{\top}(AA^{\top})^{-1} y$ is perfectly consistent with the  binary measurements, \ie $\sign{Af(y)}=y$, but is generally far from being optimal~\citep{boufounos2015quantization}.}

Here we show that it is still possible to (approximately) \red{identify the signal set and learn to reconstruct the binary measurements,} %\LJ{we must be sure of our terminology here, and align it with the title and abstract. Moreover, learning to reconstruct is not recovering $\signalset$ per se. We should explain this.} 
if the measurement operator varies across observations, \ie
\begin{equation}
    y_i = \sign{A_{g_i} x_i}
\end{equation} 
 where each signal $x_i$ is observed via one out of $\ntransf$ operators $g_i \in \{1,\dots,\ntransf\}$, and $i=1,\dots,N$. 
 This sensing assumption holds in various practical applications, where signals are observed through different operators (\eg recommendation systems access ratings about a different set of items for each user) or through an operator which changes through time (\eg a sensor that changes its calibration). Moreover, this assumption is also valid for the case where we obtain binary measurements via a single operator $A$, but the set $\signalset$ is known to be invariant to a group of invertible transformations $\{T_g\}_{g=1}^{\ntransf}$, such as translations or rotations. The invariance of $\signalset$ provides access to measurements associated with a set of (implicit) operators $\{A_g = AT_g\}_{g=1}^{\ntransf}$, as we have that
\begin{equation}
    y = \sign{AT_g T_g^{-1}x} = \sign{AT_g x'} 
\end{equation}
with $x'=T_g^{-1}x \in \signalset$ for all $g=1,\dots,\ntransf$.
This observation has been exploited to perform fully unsupervised learning on various linear inverse problems, such as magnetic resonance imaging and computed tomography~\citep{chen2021equivariant,chen2021robust,tachella2022sensing}. 
 
 %The two fundamental problems of signal and model identification from binary observations can be summarised as follows: 
%\begin{description}
%    \item[Signal Recovery] What error should we expect when estimating $x$ from the measurements $y=\sign{A_gx}$ under the constraint that $x$ belongs to a low-complexity signal set $\signalset$?
%    \item[Model Identification] How well can we estimate the set of signals $\signalset$ from the  binary measurement sets $\{\sign{A_g\signalset}\}_{g=1}^{G}$?
%\end{description}

The problem of recovering a signal from binary measurements under the assumption of a known signal set has been extensively studied in the literature~\citep{goyal1998quantized,jacques2013robust,oymak2015near}. These works provide practical bounds that characterize the recovery error as a function of the number of measurements $m$ for different classes of signal sets. \red{However, they assume that the signal set is known (or that there is enough ground-truth training data to approximate it), which is not often the case in real-world scenarios. Here we investigate the best approximation of the signal set that can be obtained from the binary observations. This approximation lets us understand how well we can learn the reconstruction function from binary data.
To the best of our knowledge, the model identification problem has not been yet addressed, and we aim to provide the first answers to this problem here.} The main contributions of this paper are:
\begin{itemize}
    \item We show that for any $G$ sensing matrices $A_1,\dots, A_{\ntransf}\in \R{m\times n}$ and any dataset size $N$, there exists a signal set whose identification error (precisely defined in \Cref{sec: model ident}) from binary measurements cannot decay faster than $\mathcal{O}(\frac{n}{m\ntransf})$ when $m$ increases.
% ###Backup of the sentence modified above###: \item We show that for any $G$ sensing matrices $A_1,\dots, A_{\ntransf}\in \R{m\times n}$ \red{and any dataset size $N$,} the signal set cannot be estimated from binary measurements up to a global identification error (precisely defined in \Cref{sec: model ident}) which decays faster than $\mathcal{O}(\frac{n}{m\ntransf})$. 

    %\LJ{How different this claim is from the linear case? From $G$ measurement vectors, we cannot recover a continuous set anyway from linear measurements. So how this contribution could be compared to the linear case?}
    \item We prove that, if each operator $A_g$, $g \in \{1,\dots,\ntransf\}$, has iid Gaussian entries (a standard construction in one-bit compressed sensing), it is possible to estimate a $k$-dimensional\footnote{The definition of dimension used in this paper is the upper box-counting dimension defined in~\Cref{sec: signal recovery}.} signal set up to a global error of $\mathcal{O}( \frac{k + n/\ntransf}{m} \log \frac{nm}{k+n/\ntransf} )$ with high probability. %, as long as the number of operators is sufficiently large, \ie $\ntransf\approx n$.
     \item We determine the \emph{sample complexity} of the related unsupervised learning problem, \ie we find that, \review{for $\ntransf$ operators with Gaussian entries,} the number of distinct binary observations for obtaining the best possible approximation of a $k$-dimensional signal set $\cl X$ \review{is $N = \mathcal{O}\big(\ntransf(\frac{m \sqrt n}{k})^{5k}\big)$ with controlled probability, which reduces to $N = \mathcal{O}\big(\ntransf(\frac{m}{k})^{k}\big)$ if $\cl X$ is a union of $k$-dimensional subspaces}. 
    \item We introduce a \red{Self-Supervised learning loss for training reconstruction networks from Binary Measurement data alone (\proposed),}  and show experimentally that the learned reconstruction function outperforms classical binary iterative hard thresholding~\citep{jacques2013robust} and performs on par with fully supervised learning on various real datasets. %\LJ{This contribution, in addition to being experimental, offers an implicit way to learn the signal set. This should be clarified in the text as well as in certain notations and in the cost functions. I propose some solutions later for that.}
\end{itemize}
A summary of the model identification bounds presented in this paper is shown in~\Cref{tab: summary}.
\begin{table}[t] 
\centering
\scalebox{0.9}{
\begin{tabular}{|l|c|c|c|c|c|}
\hline
%Lower bound inters. cells& - & $2^{n}(\frac{em\ntransf}{n})^{m\ntransf}$ & $2^{3 S k \log(\frac{m \sqrt n}{S k})}$ &  $2^{n}(\frac{em\ntransf}{n})^{m\ntransf}$   &      -              \\ \hline
&&&\\[-3.5mm]
Assumption on $\signalset \subseteq \Sp{n-1}$ & None                                              & None            & $\boxdim < k$ \\[1mm] \hline
&&&\\[-3.5mm] 
Assumption on $A_g\in \R{m\times n}$, $g \in [G]$ & rank $[A_1^\top, \ldots, A_G^\top] < n$                                              & None            & Gaussian \\[1mm] \hline
&&&\\[-3.5mm]
Identification error bounds   & $\delta>1$   & $\delta\gtrsim\frac{n}{m\ntransf}$ & $\delta \lesssim \frac{k+n/\ntransf}{m} \log \frac{nm}{k+n/\ntransf}$     \\[1mm]  \hline
        \small \red{Section}& \small \red{\Cref{subsec: lower bound}}      & \small \red{\Cref{subsec: lower bound}}   &  \small \red{\Cref{subsec: sufficient cond}}                    \\ \hline
\end{tabular}}
\caption{
%\LJ{I've modified a bit this table. Somehow, \Cref{prop: necessary multA}, \Cref{prop:  lowerbound} and \Cref{theo: onebit} do not directly provide these bounds. They still need to be combined with the context around to get these bounds. We could have to rephrase these propositions to clarify this.}
Summary of the global model identification error $\delta$ bounds presented in this paper. \review{The identification error $\delta$ corresponds to the maximal error of the optimal estimation of the signal set from binary measurement data alone (see~\Cref{def: delta}). The bounds depend on} the size of the signals $n$, the number of binary measurement operators $\ntransf$ with $m$ measurements, and the dimension of the signal set $k$. }
\label{tab: summary}
\end{table}

%The paper is organized as follows: 

\subsection*{Related Work}

\paragraph{Unsupervised learning in inverse problems.}
Despite providing very competitive results, most deep learning-based solvers require a supervised learning scenario, \ie they need measurements and signal pairs $\{(y_i,x_i)\}$, a labeled dataset, in order to learn the reconstruction function $y\mapsto x$. A first step to overcome this limitation is due to Noise2Noise~\citep{lehtinen2018noise2noise}, where the authors show that it is possible to learn from only noisy data if two noisy realizations of the same signal $\{(x_i+n_i,x_i+{n}_i')\}$ are available for training. This approach has been extended to linear inverse problems with pairs of measurements $\{(A_{g_i}x_i+n_i,A_{g_i'}x_i+{n}_i')\}$ ~\citep{yaman2020self,liu2020rare}.
The equivariant imaging framework~\citep{chen2021equivariant,chen2021robust} shows that learning the reconstruction function from unpaired measurement data $\{Ax_i+n_i\}$ of a single incomplete linear operator $A$ is possible if the signal model is invariant to a group of transformations. This approach can also be adapted to the case where the signal model is not invariant, but measurements are obtained via many different operators $\{A_{g_i}x_i+n_i\}$~\citep{tachella2022unsupervised}.
Necessary and sufficient conditions for learning in these settings are presented in~\citet{tachella2022sensing}, however under the assumption of linear observations (no quantization). Here we extend these results to the non-linear binary sensing problem with \red{an unsupervised dataset with multiple operators $\{\sign{A_{g_i}x_i}\}$ and $g_i\in \{1,\dots,\ntransf\}$, or with a single operator and a group-invariant signal set $\{\sign{Ax_i}\}$.}

\paragraph{Quantized and one-bit sensing.}
Reconstructing signals from one-bit compressive measurements is a well-studied problem~\citep{goyal1998quantized,jacques2013robust,oymak2015near,baraniuk2017onebit}, both in the (over)complete case $m\geq n$~\citep{goyal1998quantized}, and in the incomplete setting $m<n$, either under the assumption that the signals are sparse~\citep{jacques2013robust}, or more generally, that the signal set has small Gaussian width~\citep{oymak2015near}. Some of these results are summarized in~\Cref{sec: signal recovery}.
The theoretical bounds presented in this paper complement those of signal recovery bounds from quantized data, as they characterize the fundamental limitations of model identification from binary measurement data.

%The fundamental limitation of failing to learn a signal model from incomplete measurement data goes back to blind compressed sensing~\citep{gleichman2011blind}, for the specific case of models exploiting sparsity on an orthogonal dictionary. In order to learn the dictionary from compressed observations, 
%Gleichman and Eldar~\citep{gleichman2011blind} imposed additional constraints on the dictionary, while some subsequent papers~\citep{silva2011blind,aghagolzadeh2015new} removed these assumptions by proposing to use multiple operators $A_g$ as studied here. This paper can be seen as a generalization of such results to a much wider class of signal models.

\paragraph{One-bit matrix completion and dictionary learning.} 
%\JT{do we want to mention this? The requirement of noise is slightly confusing with our narrative.}
Matrix completion consists of inferring missing entries of a data matrix $Y = [y_1,\dots,y_N]$, whose columns can be seen as partial observations of signals $x_i$, \ie $y_i = \sign{A_{g_i}x_i}$ where the operators $A_{g_i}$ select a random subset of $m$ entries of the signal $x_i$. In order to recover the missing entries, it is generally assumed that the signals $x_i$ (the columns of $X=[x_1,\dots,x_N]$) belong to a $k$-dimensional subspace with $k\ll n$. \citet{davenport2014onebit} solve this learning problem via convex programming and present theoretical bounds for the reconstruction error.

\citet{zayyani2015dictionary} present an algorithm that learns a dictionary (\ie a union of $k$-dimensional subspaces) from binary data alone in the overcomplete regime $m>n$.
\citet{rencker2019sparse} presents a similar dictionary learning algorithm with convergence guarantees.
In this paper, we characterize the model identification error for the larger class of low-dimensional signal sets, which includes subspaces and the union of subspaces as special cases. \red{Moreover, we propose a self-supervised method that learns the reconstruction mapping directly, avoiding an explicit definition (\eg a dictionary) of the signal set.}
%\LJ{We should also add that, for the experimental part, we do not get an explicit regularizer adjusted by $\signalset$, but rather a self-supervised reconstruction algorithm that contains the information of both $\signalset$ and the sensing context.}

\section{Signal Recovery Preliminaries}\label{sec: signal recovery}

\begin{figure}[t]
\centering
\includegraphics[width=1\textwidth]{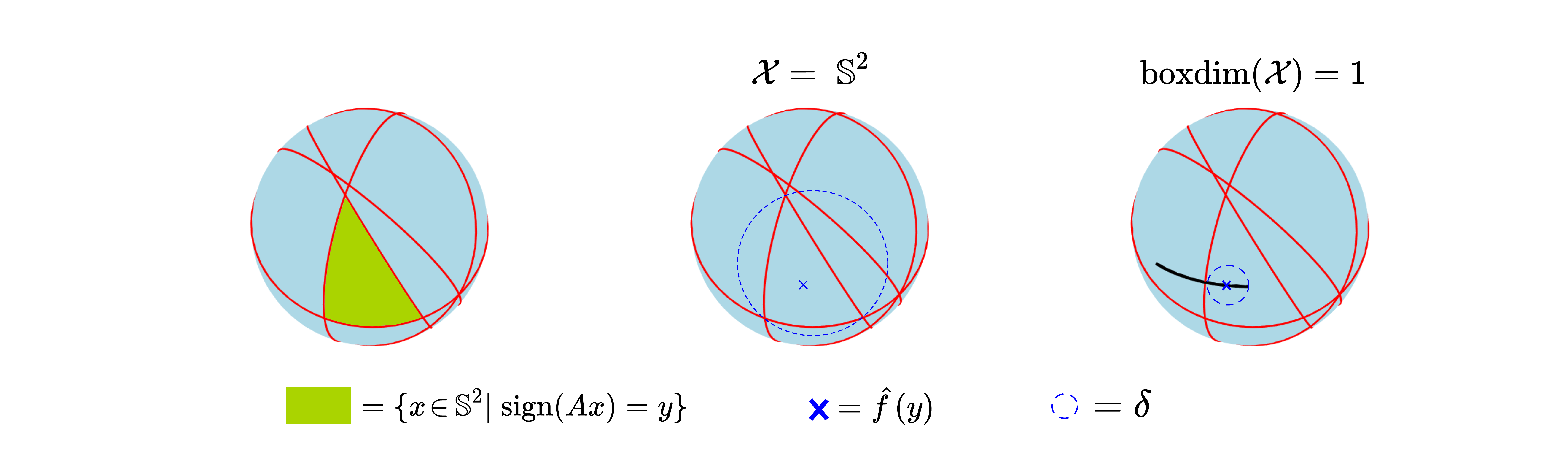}
\caption{Geometry of the 1-bit signal recovery problem with $m=5$ and $n=3$. \textbf{Left:} The binary sensing operator $\sign{A\cdot}$ defines a tessellation of the sphere into multiple \emph{consistency cells}, which are defined as all vectors $x\in \Sp{2}$ associated with the same binary code. The consistency cell associated with a given measurement $y$ is shown in green. Each red line is a great circle defined by all points of $\Sp{2}$ perpendicular to one row of $A$. 
\textbf{Middle:} If the signal set consists of all vectors in the sphere, \ie $\signalset = \Sp{2}$, the center of the cell is the optimal reconstruction $\hat{f}(y)$ (depicted with a blue cross) and the recovery error (denoted by $\delta$) is given by the radius of the cell. \textbf{Right:} If the signal set (depicted in black) occupies only a small subset of $\Sp{2}$, \ie it has a small box-counting dimension, the optimal reconstruction corresponds to the center of the intersection between the signal set and the consistency cell, and the resulting signal recovery error is smaller.}
\label{fig: recovery schematic} 
\end{figure}
We begin with some basic definitions related to the one-bit sensing problem. The diameter of a set is defined as $\diam(S) = \sup_{u,v\in S} \|u-v\|$, and the radius is defined as half the diameter. Each row $a_i\in \R{n}$ in the operator $A$ divides the unit sphere $\Sp{n-1}$ into two hemispheres, \ie  $\{x\in \Sp{n-1} : a_i^{\top}x \geq 0\}$ and $\{x\in \Sp{n-1} : a_i^{\top}x < 0\}$. Considering all rows, the operator $\sign{A\cdot}$ defines a \emph{tesselation} of $\Sp{n-1}$ into  \emph{consistency cells}, where each cell is composed of all the signals that are associated with a binary code $y$, \ie $\{ x\in \Sp{n-1}: \sign{Ax} = y\}$. The radius and number of consistency cells play an important role in the analysis of signal recovery and model identification. 
\Cref{fig: recovery schematic} illustrates the geometry of the problem for $n=3$ and $m=5$.

The problem of recovering a signal from one-bit compressed measurements with a known signal set has been well studied~\citep{goyal1998quantized,jacques2013robust,oymak2015near,baraniuk2017onebit}. These works characterize the maximum estimation error across all signals obtained by an optimal reconstruction function $\hat{f}$, \ie
\begin{equation}
\delta = \max_{x\in\signalset} \; \|x- \hat{f}( \sign{Ax})\|
\end{equation}
 as a function of the number of measurements and complexity of the signal model.
From a geometric viewpoint (see~\Cref{fig: recovery schematic}), the optimal reconstruction function with respect to the norm $\|\cdot \|$ is given by the centroid (with respect to the same norm $\|\cdot \|$) of the intersection between the consistency cell associated with the measurement $y=y(x)=\sign{Ax}$, \ie $S_y:= \{u \in \bb S^{n-1}: y=\sign{Au}\}$, and the signal set $\signalset$, \ie 
\begin{equation} \label{eq: oracle f}
    \hat{f}(y) = \text{centroid}(S_{y} \cap \signalset).
\end{equation}
while the maximum reconstruction error is given by the intersection with maximal radius, that is
\begin{equation}
   \delta = \max_{x\in\signalset} \; \text{radius}(S_{y(x)} \cap \signalset).
\end{equation}
In the overcomplete case $m>n$, assuming that all unit vectors are plausible signals, \ie $\signalset=\Sp{n-1}$, the mean reconstruction error \corr{$\delta$ is given by the consistency cell with maximal radius, which scales as $\frac{n}{m}$ (see~\Cref{prop:  lowerbound}). The optimal rate is achieved by} measurement consistent reconstruction functions, \ie those verifying $y=\sign{Af(y)}$~\citep{goyal1998quantized}.

In the incomplete case $m<n$, non-trivial signal recovery is only possible if the set of signals occupies a low-dimensional subset of the unit sphere $\Sp{n-1}$~\citep{oymak2015near}. For example, a common assumption is that $\signalset$ is the set of $k$-sparse vectors~\citep{jacques2013robust}.
In this paper, we characterize the class of low-dimensional sets using a single intuitive descriptor, the box-counting dimension.
The upper box-counting dimension~\cite[Chapter~2]{falconer2004fractal} is defined for a compact subset $S\subset\R{n}$ as
\begin{equation}
   \bdim{S} = \lim \sup_{\epsilon\to0^{+}}  \frac{\log \mathfrak{N}(S,\epsilon)}{\log 1/\epsilon}
\end{equation}
where $\mathfrak{N}(S,\epsilon)$ is the minimum number of closed balls of radius $\epsilon$ with respect to the norm $\|\cdot\|$ that are required to cover $S$.
This descriptor has been widely adopted in the inverse problems literature~\citep{puy2017recipes,tachella2022sensing}, and it captures the complexity of various popular models, such as smooth manifolds~\citep{baraniuk2009random} and union of subspaces~\citep{blumensath2009uos,baraniuk2017onebit}. For example, the set of $(k+1)$-sparse vectors with unit norm has a box-counting dimension equal to $k$.
The upper box-counting dimension is particularly useful to obtain an upper bound on the covering number of a set: if $\bdim{\signalset}<k$, there exists a set-dependent constant $\epsilon_{0} \in (0,\frac{1}{2})$ for which
\begin{equation}
 \mathfrak{N}(\signalset,\epsilon)\leq \epsilon^{-k}
\end{equation}
holds for all $\epsilon\leq \epsilon_0$~\citep{puy2017recipes}. The following theorem (proved in \Cref{app: signal recovery}) exploits this fact to provide a bound on the number of measurements needed for recovering a signal with an error smaller than $\delta$ from generic binary observations.

\begin{theorem}\label{theo: signal recov boxdim}
Let $A$ be a matrix with iid entries sampled from a standard Gaussian distribution and assume that $\bdim{\signalset}<k$, \red{such that $ \mathfrak{N}(\signalset,\epsilon)\leq \epsilon^{-k}$ for all $\epsilon<\epsilon_0$ with $\epsilon_0\in (0,\frac{1}{2})$.} \review{For $\delta\leq \min \{ \review{30}\sqrt{n}\epsilon_0,\frac{1}{2}\}$}, if the number of measurements verifies
\begin{equation}
\label{eq:thm-consist-cond}
m \geq  \tfrac{\review{4}}{\delta} \big(2k\log\tfrac{\review{30}\sqrt{n}}{\delta} + \log\tfrac{1}{\xi} \big)
\end{equation}
then for all $x,s\in \signalset$, we have that 
\begin{equation}
\label{eq:thm-consist-eqt}
   \sign{Ax} \review{=} \sign{As} \implies \|x-s\| < \delta 
\end{equation}
with probability greater than $1-\xi$.
\end{theorem}
\noindent This result extends Theorem 2 in~\citet{jacques2013robust}, which holds for $k$-sparse sets only, to general low-dimensional sets and is included in \Cref{app: signal recovery}. 
For example, if $\signalset$ is the intersection of $L$ $(s+1)$-dimensional subspaces with the unit sphere, %$N(\signalset,\delta)\leq L(\delta/3)^{-k}$ and 
\Cref{theo: signal recov boxdim} holds with constant $\epsilon_0 = (3^{s}L)^{-\frac{1}{k-s}}$ and $k>s$~\cite[Chapter~4.2]{vershynin2018high}. %\LJ{reference?}
This theorem tells us that we can recover sparse signals from binary measurements up to an error of $$\mathcal{O}(\tfrac{k}{m}\log \tfrac{nm}{k})$$ which is sharp, up to the logarithmic factor~\citep{jacques2013robust}. Oymak and Recht~\citep{oymak2015near} present a similar result, stated in terms of the Gaussian width\footnote{The Gaussian width of a set $S$ is defined as $\mathbb{E}_s\{ \sup_{x\in S} x^{\top}s \}$ where $s$ is distributed as a standard Gaussian vector.} of the signal set instead of the box-counting dimension.
%\JT{Do we want to include our own general recovery theorem here? I think we should get an error of order $\mathcal{O}\left(\frac{k+kc_{\signalset}}{m}\log\frac{m\sqrt{n}}{k}\right)$}
 \begin{figure}[t]
\centering
\includegraphics[width=1\textwidth]{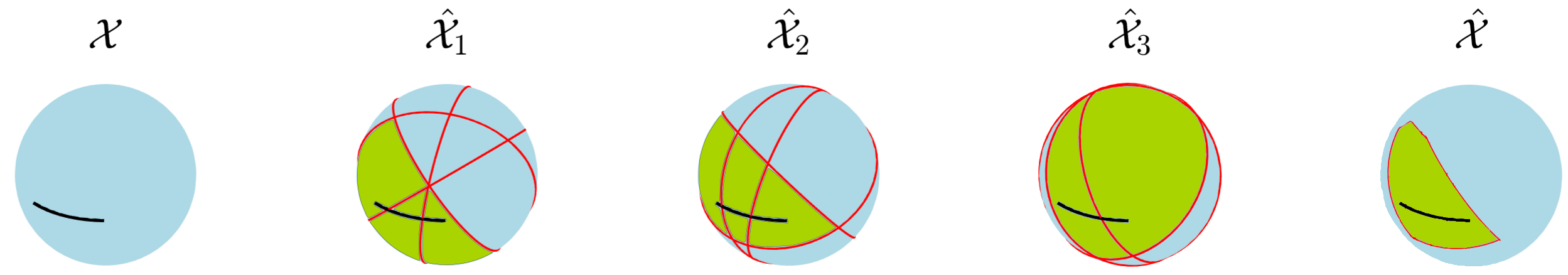}
\caption{%\LJ{Note that $\hat{\signalset}_i$ can be non-convex, in the same way than $\signalset_{\rm oracle}$ is in \Cref{fig:oracle}}
Illustration of the model identification problem from binary measurements with $n=3$, $m=4$, and $G=3$. A signal set with box-counting dimension 1 is depicted in black. The red lines define the frontiers of the consistency cells associated with operators $A_1,\dots, A_3$. \textbf{From left to right:} The signal set, the estimation of the signal set associated with $A_1,\dots,A_3$ and the overall estimate $\hat{\signalset}$.} %, \ie each red line is a great circle defined by all points of $\Sp{2}$ perpendicular to one row of a given operator.} %\LJ{As said on page 1, this is why I had to define the concept of "cell"; it's obvious for us, not necessarily for the reader}
\label{fig:illustration} 
\end{figure}

\section{Model Identification from Binary Observations} \label{sec: model ident}

In this section, we study how well we can identify the signal set from binary measurement data associated with $\ntransf$ different measurement operators $A_1,\dots,A_{\ntransf}\in \R{m\times n}$. We focus on the problem of identifying the set $\signalset$ from the binary sets $\{\sign{A_g\signalset}\}_{g=1}^{\ntransf}$. In practice, we observe a subset of each binary set $\sign{A_g\signalset}$, \corr{however, in \Cref{subsec: complexity} we show that} the number of elements in each of these sets is controlled by the box-counting dimension of $\signalset$, which is typically low in real-world settings~\citep{hein2005intrinsic}.

We start by analyzing how the different operators provide us with information about $\signalset$. 
Each forward operator $A_g$ constrains the signal space by the following  set \begin{equation} \label{eq: constraint Ag}
    \hat{\signalset}_g = \{v \in \Sp{n-1} : \; \exists x_g\in \signalset, \;  \sign{A_gv} = \sign{A_g x_g}\}.
\end{equation}
Each set $\hat{\signalset}_g$ is thus composed of all unit vectors $v$ that are \emph{consistent} with at least one point $x_g$ of $\signalset$ according to the binary mapping $\sign{A_g\cdot}$. %, \ie $\sign{A_gv} = \sign{A_g x_g} =: y_g$. 
%Said differently, defining the concistency cell $\hat{\cl X}_{y_g,g} \subset \bb S^{n-1}$ of a given point $x \in \signalset$ as the set of all unit vectors $v$ \emph{consistent} with $x$ according to $\sign{A_g\cdot}$, then $\hat{\signalset}_g$ is the union $\cup_{x \in \signalset} \,\hat{\cl X}_{y_g,g}$ of all consistency cells defined over all points of $\signalset$. 
We thus conclude that $\hat{\signalset}_g$ is essentially a \emph{dilation} of $\signalset$---and we clearly have $\signalset \subset \hat{\signalset}_g$---whose extension is locally determined by specific cells of $\sign{A_g\cdot}$.  A three-dimensional example with $m=4$ measurements and $\ntransf=3$ operators is presented in~\Cref{fig:illustration}. Note that, for a given binary mapping $\sign{A_g\cdot}$, each cell is characterized by one binary vector in the range of this mapping, so that, as shown in this figure, all cells provide a different tesselation of $\Sp{n-1}$ whose size and dimension will play an important role in our analysis. %\LJ{The purpose of this new paragraph is to introduce the concepts of consistency and cells, which are helpful later and for \Cref{fig:illustration}.}

Since each $\hat{\signalset}_g$ is a dilation of $\signalset$, we can infer the signal set from the following intersection
\begin{equation}
    \hat{\signalset} := \bigcap_{g=1}^{\ntransf} \hat{\signalset}_g, 
\end{equation}
which can be expressed concisely as
\begin{equation} \label{eq:inferred set}
  \hat{\signalset} =  \left\{ v\in \Sp{n-1} :  \; \exists x_1,\dots,x_{\ntransf}\in \signalset, \; \;   \sign{A_gv} = \sign{A_g x_g}, \; \forall g =1,\dots, \ntransf  \right\}.
\end{equation}

Due to the binary quantization, the inferred set will be larger than the true set, \ie $\signalset \subset \hat{\signalset}$. However, we will show that it is possible to learn a slightly \emph{larger} signal set, defined in terms of a global identification error $\delta>0$, \ie the open $\delta$-\emph{tube}
\begin{equation}
 \review{\signalset_\delta = \{ v \in \Sp{n-1} :\;  \;\inf_{x \in \signalset}\| x- v \| < \delta\}}
\end{equation}
such that the inferred set is contained in it, \ie $\hat{\signalset} \subset \signalset_\delta$. %\LJ{I guess you need this tube to be open for \Cref{prop: rotation} below, right?}. 
\review{We define the model identification error as the smallest $\delta$ such that $\hat{\signalset} \subset \signalset_\delta$ holds:
\begin{definition}[Model identification error]\label{def: delta}
    The identification error of a signal set $\signalset\subset \Sp{n-1}$ from binary sets $\{ \sign{A_g\signalset} \}_{g=1}^{\ntransf}$ is defined as $\min \{\delta \geq 0 : \hat{\signalset} \subseteq \signalset_{\delta} \}$.
\end{definition}
}

For our developments to be valid, we will further assume that $\signalset$ is not too dense over $\bb S^{n-1}$ so that two tubes of $\signalset$ with two distinct radii are distinct.
\begin{assumption}
\label{ass:not-too-dense-X}
\red{The set $\signalset$ is closed} and there exists a maximal radius $0< \delta_0 < 2$ for which $\signalset_{\delta} \subsetneq \signalset_{\delta_0}$ for any $0<\delta<\delta_0$.
\end{assumption}
This assumption amounts to saying that there exists at least one open ball \corr{in $\bb S^{n-1}$} that does not belong to $\cl X \subset \bb S^{n-1}$. For instance, $\signalset = \bb S^{n-1}$ does not verify this assumption, and $\signalset = \bb S^{n-1} \cap \{x \in \bb R^n: x_1 \geq 0\}$ verifies it for $\delta_0 \leq \sqrt{2}$ since $\signalset_{\delta} = \bb S^{n-1}$ for any $\delta \geq \sqrt{2}$. 
 The next subsections provide lower and upper bounds for $\delta$. %with high probability with respect to a random sample of a set of operators $A_1,\dots, A_{\ntransf}$. 

\subsection{A Lower Bound on the Identification Error} \label{subsec: lower bound}

We first aim to find a lower bound on the best $\delta$ achievable via the following oracle argument: if we had oracle access to $\ntransf$ measurements of each point $x$ in $\signalset$ through each of the $\ntransf$ different operators, we could stack them together to obtain a larger measurement operator, defined as 
\begin{equation} \label{eq: oracle}
    \begin{bmatrix}
    y_1 \\
    \vdots \\
    y_{\ntransf}
    \end{bmatrix} = \sign{\concatA x}  \text{  with  }
    \concatA  =  \begin{bmatrix}
    A_1 \\
    \vdots \\
    A_{\ntransf}
    \end{bmatrix} \in \R{m\ntransf\times n}.
\end{equation}
This oracle measurement operator  provides a  refined approximation of the signal set, specified as 
\begin{equation}
\label{eq: constraint oracle}
    \hat{\signalset}_{\text{oracle}} = \{v \in \Sp{n-1} : \; \exists x\in \signalset, \;  \sign{ \concatA v} = \sign{\concatA x}\},
\end{equation}
which is again a dilation of $\signalset$.
%What is the best approximation of $\signalset$ we can obtain from the oracle binary measurement set $\sign{\concatA \signalset}$? 

\begin{figure}[t]
\centering
\includegraphics[width=.7\textwidth]{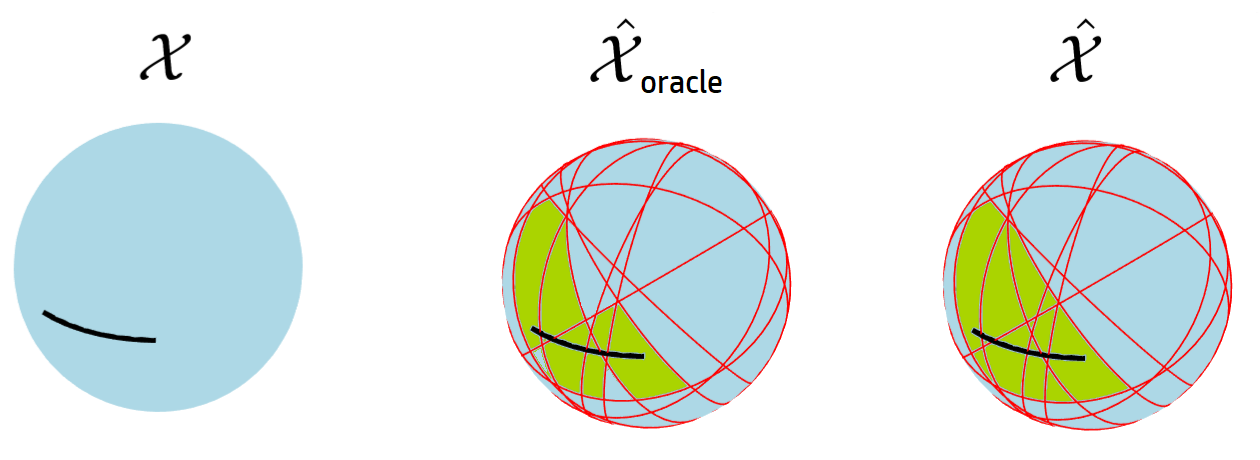}
\caption{Illustration of the oracle argument in the example of~\Cref{fig:illustration}. \textbf{Left:} The signal set $\signalset\subset \Sp{2}$ is depicted in black. \textbf{Middle:} Cells intersected by the oracle system are indicated in green.  \textbf{Right:} The identified set $\hat{\signalset}$ is indicated in green, and \emph{is  larger} than the oracle counterpart.}
\label{fig:oracle} 
\end{figure}

\Cref{fig:oracle} shows an example with the oracle set $\hat{\signalset}_{\text{oracle}}$, which provides a better (or equal) approximation of the signal set than~\Cref{eq:inferred set}, due to the fact that $\signalset\subset \hat{\signalset}_{\text{oracle}}\subseteq \hat{\signalset}$ by the construction of these sets.
 As the oracle estimate is composed of the cells associated with $\sign{\concatA \cdot}$ which are intersected by the signal set, the oracle approximation error depends on the diameter of the intersected cells. Given a certain oracle tesselation of $\Sp{n-1}$, the worst estimate of $\signalset$ is obtained when it intersects the largest cells in the tessellation.  The following proposition formalizes the intuition that the maximum consistency cell diameter---\ie the greatest distance separating two binary consistent vectors of $\signalset$ according to $\concatA $---serves as a lower bound on the model identification error $\delta$.

%\LJ{For \Cref{prop: rotation}, the rational behind the rotation trick should be explained. Moreover, for the proof to work, we must impose $\signalset$ close. To better define $\delta_0$ below, I've rewritten the proposition below and defined \Cref{ass:not-too-dense-X} above.}

\begin{proposition}\label{prop: rotation}
Given $\concatA \in\R{ m\ntransf \times n}$, for any set $\cl X \subset \bb S^{n-1}$ respecting \Cref{ass:not-too-dense-X} with $0<\delta_0 <2$, there exists a rotation matrix $R\in SO(n)$ such that the rotated set 
    \begin{equation}
        \signalset' = \{v\in \Sp{n-1} : v = Rx,\; x\in \signalset \} = R \signalset
    \end{equation}
    verifies $\hat{\signalset}_\text{oracle}' \not \subset \signalset'_{\delta}$ for any $\delta<\min\{d,\delta_0\}$ where $0<d<2$ is the largest cell diameter of the tesselation induced by $\sign{\concatA \cdot}$.
\end{proposition}
\begin{proof}
    Given $\delta < \delta_0$, the proof consists in choosing an appropriate rotation matrix, such that we can find a point $v$ which belongs to the oracle estimate $\hat{\signalset}_\text{oracle}'$ of the rotated set $\signalset'$, but doesn't belong to the $\delta$-tube $\signalset_{\delta}'$ of this set.  
   From \Cref{ass:not-too-dense-X} and since the $\delta$-tube $\signalset_{\delta}$ is open, there exists $x\in\signalset$ and $v \not\in \signalset_{\delta}$ such that $\|x-v\|=\delta$ 
   %\LJ{This is where we need $\signalset$ close}. 
   Let $S$ denote the largest cell in the tesselation of $\Sp{n-1}$ induced by $\sign{\concatA \cdot}$, such that $d=\text{diam}(S)$. If $\delta<d$, we can always pick a rotation $R\in SO(n)$ such that both $x' = R x$ and $v' = R v$ belong to $S$.  As $x'\in S$, $\signalset'$ intersects $S$ and we have that $ S\subseteq \hat{\signalset}_\text{oracle}'$, and thus that $v'\in \hat{\signalset}_\text{oracle}'$.
\end{proof}

\review{In words, \Cref{prop: rotation} shows that we can rotate any signal set $\signalset$ such that it intersects the largest consistency cell in the tesselation, obtaining a model identification error that is proportional to the maximum cell diameter. The rotation is used to remove the best-case scenarios where the signal set only intersects consistency cells that are smaller than the largest one.}

In the rest of this subsection, we focus on bounding the maximum cell diameter, as it is directly related to the model identification error through \Cref{prop: rotation}. We start with the following proposition which shows that, if the stacked matrix is rank-deficient, \review{all} %there exist 
cells \review{have} the maximum possible diameter.

\begin{proposition}%[Theorem 1 in~\citep{chen2021equivariant}] 
\label{prop: necessary multA}
Consider the tessellation defined by $\sign{\concatA \cdot}$   
with $\concatA \in\R{m\ntransf\times n}$. If \begin{equation}\label{eq: rank condition}
    \rk{\concatA } < n
\end{equation} 
\review{all the cells} %there are cells 
in the tessellation \review{have} a diameter equal to $2$.
\end{proposition}
\begin{proof}
If $\concatA $ has a rank smaller than $n$, it has a non-trivial nullspace. Let $v\in \Sp{n-1}$ be \review{an element in} the nullspace with unit norm.
Consider a cell associated with the code $\sign{\concatA x}$ for some $x\in\R{n}$ inside the \review{complement of this nullspace (\ie in the range of $\concatA ^{\top}$)}. The points $\frac{x+v}{\|x+v\|},\frac{x-v}{\|x-v\|}\in \Sp{n-1}$ belong to this cell since they share the same code. As $\|x\pm v\| = \sqrt{\|v\|^2+\|x\|^2}$ due to orthogonality, the distance between these two points is 
\begin{equation}
   \frac{2\|v\|}{\sqrt{\|v\|^2+\|x\|^2}} =  \frac{2}{\sqrt{1+\|x\|^2}}
\end{equation}
which tends to $2$ as $\|x\|$ goes to zero, without modifying the cell code $\sign{\concatA x}$.
\end{proof}
\noindent %It provides a practical necessary condition for model identification, \ie that we have at least
%$$m > \frac{n}{\ntransf}$$
%measurements per operator.
\review{
This result provides a practical necessary condition for model identification, which is summarized in the following corollary:
\begin{corollary}\label{cor: necessary condition mG} A necessary condition for the tesselation defined by  $\sign{\concatA \cdot}$ to have consistency cells with a diameter smaller than 2 is that there are at least
$$m \geq {n}/{\ntransf}$$
measurements per operator.
\end{corollary}
This proposition tells us that $n/\ntransf$ measurements are necessary in order to obtain non-trivial cell diameters, and thus to obtain a non-trivial estimation of $\signalset$. 
}

Moreover, in practice, it is possible to compute the rank of the stacked matrix $\concatA $ via numerical approximations.
The following theorem provides a more refined characterization of the oracle error for $m\geq n/\ntransf$:
%\begin{theorem}
%\label{prop:  lowerbound}
%Let the entries of $A_1,\dots,A_{\ntransf} \in \R{m\times n}$ be sampled from a standard Gaussian distribution.
%The  signal set estimated from the measurement sets $\{\sign{A_g\signalset} \}_{g=1}^{\ntransf}$, as defined in \Cref{eq:inferred set}, verifies $\hat{\signalset}\not\subset\signalset_{\delta}$ for any
%$$\delta  \leq  \min\{ \delta_0, \frac{\log (\frac{1}{1-\xi})}{2m\ntransf \sqrt{n\frac{\pi}{2}}} \}$$
% with probability at least $1-\xi$, where $\delta_0\leq 2$ is a constant such that $\signalset_{\delta} \neq \Sp{n-1}$ for any $\delta<\delta_0$.
%\end{theorem}

\begin{proposition}
\label{prop:  lowerbound}
Consider the tessellation defined by $\sign{\concatA \cdot}$   
with $\concatA \in\R{m\ntransf\times n}$. The largest cell in the tessellation 
has a diameter of size at least $\frac{2}{3}\frac{n}{m\ntransf}$.
\end{proposition}

\begin{proof}
According to~\citet[Theorem A.7]{thao1996lower}, the maximum number of cells $C_{\concatA }$ induced by a tessellation defined by $\sign{\concatA \cdot}$ 
with $\concatA \in\R{m\ntransf\times n}$ can be upper bounded as
$$
    C_{\concatA } \leq \binom{m\ntransf}{n}2^{n}.
    $$
As $\binom{m\ntransf}{n}\leq (\frac{e m\ntransf}{n})^{n}$, we have that $C_{\concatA }\leq (\frac{2e m\ntransf}{n})^{n}$. 
We can inscribe all cells into spherical caps $S_i$\footnote{A spherical cap of radius $r$ around a point $v\in\Sp{n-1}$ is defined as $\{x\in \Sp{n-1}: \|x-v\|<r\}$.} of radius $\delta/2$, where $\delta$ is the maximum cell diameter. 
As shown in~\cite[Lemma 2.3]{ball1997elementary}, a spherical cap of radius $\delta/2$ has measure bounded by $\sigma(S_i) \leq (\frac{\delta}{4})^{n-1}\sigma_{n-1}$ where $\sigma_{n-1}$ is the measure of $\Sp{n-1}$. %\JT{fix this}
Since the tessellation covers the unit sphere $\Sp{n-1}$, we have that $\Sp{n-1} \subseteq \cup_{i=1}^{C_{\concatA }} S_i$ and thus
$$
\ts  \sum_{i=1}^{C_{\concatA }} \sigma( S_i) \geq \sigma_{n-1}\ \Rightarrow\ (\frac{2e m\ntransf}{n})^{n}(\frac{\delta}{4})^{n}\sigma_{n-1} \geq \sigma_{n-1}\ \Rightarrow\ \delta \geq \frac{2}{3}\frac{n}{m\ntransf}.
$$
\end{proof}
\review{\paragraph{Remark} 
The upper bound in this proposition is tight in the sense that there exist matrices that attain this bound. As a special case of \Cref{theo: signal recov boxdim}  with $mG > n$ measurements and the box-counting dimension of $\signalset$ set to $n$, for an $mG \times n$ Gaussian random matrix $\bar A$, and by choosing the minimal number of required measurements in the condition of this theorem, we have with high probability that the largest cell of $\sign{\bar A \cdot}$ has a diameter that decays like $\mathcal{O}(\frac{n}{mG})$ up to log factors. 
By a standard boosting argument\footnote{\review{Assuming a Gaussian matrix will fail to have the desired cell size with probability at most $\xi<1$, the probability that $1$ out of $r$ independent Gaussian matrices will fail to have this property is at least $1-\xi^{r}$, which can be made arbitrarily high by increasing $r$. }}, it thus means that there exists a $mG \times n$ matrix $\bar A$ with the same consistency cell diameter decay.}

As stated in the following corollary, \Cref{prop:  lowerbound} shows that the model identification error cannot decrease faster with the number of measurements and operators than $\mathcal{O}(\frac{n}{m\ntransf})$, since the largest cell in any oracle tesselation has a diameter of at least $\frac{2}{3}\frac{n}{m\ntransf}$.

\begin{corollary}
    \label{cor:lower-bound-identif-error}
    Given $G$ operators $A_1, \ldots, A_G \in \bb R^{m \times n}$, and any set $\cl X \subset \bb S^{n-1}$ verifying \Cref{ass:not-too-dense-X} with $0<\delta_0<2$, for any $0<\delta<\min(\delta_0, \frac{2}{3} \frac{n}{mG})$, there exists a rotation $R$ such that the inferred signal set $\hat{\cl X}' $ of $\cl X' = R \cl X$ is not included in $\cl X'_\delta$, \ie $\hat{\cl X}' \not\subset \cl X'_\delta$.
\end{corollary}

\begin{proof}
If $\cl X \subset \bb S^{n-1}$ respects \Cref{ass:not-too-dense-X} with $0<\delta_0<2$, and $\hat{\cl X}_{\rm oracle}$ and $\hat{\cl X}$ are the oracle set associated with $\concatA$ and the inferred set of $\cl X'$, respectively, then, as derived previously, we know that $\cl X \subset \hat{\cl X}_{\rm oracle} \subset \hat{\cl X}$. According to \Cref{prop: rotation} and \Cref{prop:  lowerbound}, there exists a rotation $R$ such that 
$\hat{\cl X}'_{\rm oracle} = R\hat{\cl X}_{\rm oracle} \not\subset \cl X'_\delta = R\cl X_\delta$ for $0<\delta<\min(\delta_0, \frac{2}{3} \frac{n}{mG})$. Therefore, from the inclusion above, we thus see that there exists $\hat{\cl X}' = R\hat{\cl X} \not\subset \cl X'_\delta$.
\end{proof}

%The constant $\delta_0$ is handles degenerate cases where $\signalset$ is too big, such as $\signalset=\Sp{n-1}$ (and thus $\delta_0=0$), and the error is trivially low since $\hat{\signalset}=\signalset$. 

\review{
At this stage, it is natural to ask if the condition in~\Cref{cor: necessary condition mG} is sufficient to upper bound the model identification error $\delta$. The answer is negative since, for certain families of operators, the maximum consistency cell associated with $\sign{\bar{A}\cdot}$ does not decrease with the number of measurements $m$ or operators $\ntransf$, as illustrated by the following example inspired by a case considered in \cite[Sec. 1.1]{plan2013one}.
\begin{example}
    Consider the operators with binary\footnote{\review{This example can be generalized to operators with entries belonging to a discrete set $Q$, and show that there exist cells with diameter equal to $\sqrt{\Delta}$ where $\Delta$ is the minimum distance between two elements in $Q$.}} entries $A_1,\dots,A_{\ntransf}\in \{-1,1\}^{m\times n}$. Let $x_{\lambda} = e_1 + \lambda e_2 \in \R{n}$ where $e_i \in \Sp{n-1}$ is the $i$th canonical vector and $\lambda$ is a scalar. Due to the quantization of the operators, we have that $\sign{\bar{A}x_{\lambda}} = \sign{\bar{A}x_{\lambda'}}$ for any value of $\lambda,\lambda' \in (-1,1)$. Thus, there exists a cell in the oracle tesselation associated with $\sign{\bar{A}\cdot}$ which contains the set of points $\{\frac{x_{\lambda}}{\|x_{\lambda}\|}\}_{\lambda \in (-1,1)}$ and thus has a diameter equal to $\sqrt{2}$, independently of the values of $m$ and $\ntransf$.
\end{example}
In the next subsection, we obtain an upper bound on the model identification error which overcomes this pathological example by sampling the operators from a continuous random distribution.
}

\subsection{A Sufficient Condition for Model Identification} \label{subsec: sufficient cond}

We now seek a sufficient condition on the number of measurements per operator that guarantees the identification of $\signalset$ up to a global error of $\delta$. As with the sufficient conditions ensuring signal recovery (see \Cref{sec: signal recovery}), we assume that $\signalset$ is low-dimensional to provide a bound that holds with high probability if the entries of the operators are sampled from a Gaussian distribution.
\begin{theorem} \label{theo: onebit}
Given the operators $A_1,\dots,A_{\ntransf} \in \R{m\times n}$ with entries \iid as a standard Gaussian distribution, a low-dimensional signal set $\cl X$, with $\bdim{\signalset}<k$, such that $\mathfrak{N}(\signalset,\epsilon)\leq \epsilon^{-k}$ for all $\epsilon<\epsilon_0$ with $\epsilon_0\in (0,\frac{1}{2})$. \review{For $0<\delta\leq \min \{ 18\sqrt{n}\epsilon_0,1\}$} and some failure probability $0 < \xi < 1$, if the number of measurements per operator verifies
\begin{equation} \label{eq: m bound}
    \textstyle m \geq \frac{4}{\delta}\, \big[(k+\frac{n}{\ntransf}) \log \frac{\review{54}\sqrt{n}}{\delta} + \frac{1}{\ntransf} \log \frac{1}{\xi} \big] %+ \frac{n}{\ntransf}\log 3\big] 
\end{equation}
then with probability at least $1-\xi$, we have that $\hat{\signalset}\subseteq \signalset_{\delta}$. %\JT{check $\delta_0$ in the proof.}
\end{theorem}

\noindent The proof is included in \Cref{app: model ident}. \Cref{theo: onebit} provides a bound on $\delta$, \ie how precisely we can characterise the signal set $\signalset$, which we can compare with the lower bound in~\Cref{prop:  lowerbound}. From~\Cref{eq: m bound} we have that (see \Cref{app: model ident} for a detailed derivation),
\begin{equation} \label{eq: main bound}
 \textstyle \delta=\mathcal{O}( \frac{k+n/\ntransf}{m} \log \frac{nm}{k+n/\ntransf} ).
\end{equation}
The bound in~\Cref{eq: main bound} is consistent with existing model identification bounds in the linear setting~\citep{tachella2022sensing}, which require  $m>k+n/\ntransf$ measurements per operator for uniquely identifying the signal set.

\subsection{Learning to Reconstruct}

The best reconstruction function $\hat{f}$ that can be learned from binary measurements alone can be defined as a function of the identified set $\hat{\signalset}$, as defined in~\Cref{eq: oracle f}:
\begin{equation}\label{eq: unsup oracle}
   \hat{f}(y) = \text{centroid}(S_{y} \cap \hat{\signalset})
\end{equation}
for a binary input $y$ with associated consistency cell $S_{y}=\{v\in\Sp{n-1} : \; \sign{Av}=y\}$. The reconstruction error of $\hat{f}$ is lower bounded by the radius of the set $S_{y} \cap \hat{\signalset}$. 
\review{This error must be larger than the error of a reconstruction function that has full knowledge about the signal set $\signalset$. Intuitively, if we have a large model identification error, $\hat{\signalset}$ will be a bad approximation of $\signalset$ and thus $\hat{f}$ will obtain large reconstruction errors. The following proposition formalizes this intuition, showing that the reconstruction error of $\hat{f}$ is lower bounded by the model identification error. 

\begin{proposition}\label{prop: reconstruction unsup f}
    Given $G$ operators $A_1, \ldots, A_G \in \bb R^{m \times n}$ and a set $\signalset \subset \Sp{n-1}$ with model identification error equal to $\delta$, there exist points $x_g\in\signalset$ for $g=1,\dots,\ntransf$ such that the reconstruction error is
    $$
    \|\hat{f}\left( \sign{A_gx_g}\right) -x_g \| \geq \delta/2,
    $$
    where $\hat{f}$ is the optimal reconstruction function that can be learned from the measurement data $\{ \sign{A_g\signalset}\}_{g=1}^{\ntransf}$, as defined in \Cref{eq: unsup oracle}.
\end{proposition}
\begin{proof}
 Following \Cref{def: delta} of model identification error, there exists a point $\hat{x}\in\hat{\signalset}$ such that $\|x-\hat{x}\| \geq \delta$ for all $x\in\signalset$. According to the construction of the inferred set $\hat{\signalset}$ in \Cref{eq:inferred set}, there exist some $x_1,\dots,x_{\ntransf}\in\signalset$ such that $\sign{A_g\hat{x}}=\sign{A_gx_g}$ for all $g=1,\dots,\ntransf$. Therefore, for any $g\in\{1,\dots,\ntransf\}$, the diameter of the set $S_{\sign{A_gx_g}} \cap \hat{\signalset}$ is at least  $\|x_g-\hat{x}\|$ since both $x_g$ and $\hat{x}$ belong to this set. As the optimal reconstruction function outputs the centroid of the set (as defined in~\Cref{eq: unsup oracle}), the reconstruction error of the point $x_g$ is at least $\|x_g-\hat{x}\|/2 \geq \delta/2$.
\end{proof}

Therefore, we can use the results on model identification developed in \Cref{subsec: lower bound} to lower bound the reconstruction error for the case where the function is learned from measurement data only. In particular, combining this result with \Cref{cor:lower-bound-identif-error}, we obtain that the (worst-case) reconstruction error should be larger than $\frac{1}{3}\frac{n}{m\ntransf}$. It is worth noting that this result also holds for the case where we have a single operator and group invariance, \ie when $A_g=AT_g$ for $g=1,\dots,\ntransf$.
}

\review{An upper bound on the reconstruction error is harder to obtain. } Unfortunately, \Cref{theo: signal recov boxdim,theo: onebit} do not automatically translate into a bound on the optimal reconstruction error of the reconstruction function defined in \eqref{eq: unsup oracle}.  \Cref{theo: onebit} implies that the optimal unsupervised reconstruction $\hat{f}(\sign{A_\review{g}x})$ is at most $\mathcal{O}( \frac{k+n/\ntransf}{m} \log \frac{nm}{k+n/\ntransf} )$ away from the signal set $\signalset$, but does not guarantee that it is close to $x$. Nonetheless, we conjecture that this rate holds with high probability if the operators follow a Gaussian distribution:
\begin{conjecture} \label{conj: unsup recov}
Given binary measurements from the operators $A_1,\dots, A_{\ntransf} \in \R{m\times n}$ with entries \iid from a standard Gaussian distribution, the optimal reconstruction function defined in \eqref{eq: unsup oracle}
has a maximal reconstruction error that is upper bounded as $\mathcal{O}(\frac{k+n/\ntransf}{m} \log \frac{nm}{k+n/\ntransf})$
with high probability.
\end{conjecture}
\Cref{conj: unsup recov} hypothesizes that the optimal unsupervised reconstruction function should obtain a similar performance than the supervised one, \ie $\mathcal{O}(\frac{k}{m} \log \frac{nm}{k})$ shown in \Cref{theo: signal recov boxdim}, if the number of operators is sufficiently large, \ie $\ntransf > n/k$. In the experiments in \Cref{sec: experiments}, we provide empirical evidence that supports this hypothesis.

\subsection{Sample complexity}\label{subsec: complexity}

We end our theoretical analysis of the unsupervised learning problem by bounding its \emph{sample complexity}, \ie we bound the number $N$ of \emph{distinct} binary measurement vectors $\{y_i\}_{i=1}^N$ that must be acquired for obtaining the best approximation of the signal set $\cl X$ from binary data. %In this subsection, we aim to upper bound how many (different) measurement vectors we need to observe to fully characterize the set $\hat{\signalset}$ defined in the previous section.

Since we observe binary vectors $y\in \{ \pm 1\}^{m}$, there is a limited number of different binary observations.
%\footnote{Note that observing multiple times the same binary measurement provides information about the signal distribution, but it does not provide additional information about the support $\signalset$ which is the main focus of this paper. \LJ{I guess this sentence will be unclear for most readers}}.
We could naively expect to observe up to $2^{m}$ different vectors per measurement operator (\ie all possible binary codes with $m$ bits), requiring at most $N\leq \ntransf2^{m}$ samples to fully characterize the best approximation of the signal set $\hat{\signalset}$ defined in~\Cref{eq:inferred set}.
Fortunately, as already exploited in the proof of \Cref{prop:  lowerbound}, this upper bound can be significantly reduced if the signal set has a low box-counting dimension, as not all cells in the tessellation will be intersected by the signal set (see~\Cref{fig:illustration}). We can thus obtain a better upper bound by counting the number of intersected cells, denoted as $|\sign{A\signalset}|$. 
 
If $\signalset$ is the intersection of a single $k$-dimensional subspace with the unit sphere, \cite[Theorem A.7]{thao1996lower} tells us that, for any matrix $A\in\R{m\times n}$ \review{with $m \geq k$}, there are $|\sign{A\signalset}|\leq 2^{k}\binom{m}{k}$ intersected cells. More generally, if $\signalset$ is a union of $L$ subspaces, we have $|\sign{A\signalset}|\leq L 2^{k}\binom{m}{k}$. Thus, using the fact that $\binom{m}{k}\leq \left(\frac{3m}{k}\right)^k$, from the $G$ measurement operators, we can observe up to 
\begin{equation} \label{eq: n uos}
\textstyle N \leq  \ntransf L (\frac{6 m}{k})^{k}
\end{equation}
different measurement vectors. However, this result only holds for a union of subspaces having each dimension $k$. The following theorem extends this result to more general low-dimensional sets with small upper box-counting dimension.
\review{\begin{theorem}
\label{thm:max-bits-gauss}
Let the entries of $A\in \R{m\times n}$ be sampled from a standard Gaussian distribution, and let $\signalset\subseteq \R{n}$ with $\bdim{\signalset}<k$. If $k/(m \sqrt n) < \min(\epsilon_0,1/2)$, then, in expectation, the cardinality of the measurement set is bounded as   
\begin{equation}
\label{eq:expected-number-cells-k-dim-space}
\textstyle \bb E |\sign{A \signalset}| \leq \big(\frac{e m \sqrt{n}}{k}\big)^{k}.
\end{equation}
Moreover, given a failure probability $0 < \xi < 1$, if  
$2k/(m \sqrt{n}) \leq \min(\epsilon_0,1/2)$, then, with probability $1 - \xi$, we have 
\begin{equation}
\label{eq:proba-bound-number-cells-k-dim-space}
\textstyle |\sign{A \signalset}| \leq (\frac{1}{\xi})^4 \big(\frac{3 m \sqrt{n}}{5k}\big)^{5k}.
\end{equation}
\end{theorem}}%
%\LJ{I have applied the change $S \to s$ below, as $S$ is usually used for sets.}
% \begin{theorem}\label{thm:max-bits-gauss}
% Let the entries of $A\in \R{m\times n}$ be sampled from a standard Gaussian distribution, and let $\signalset\subseteq \R{n}$ with $\bdim{\signalset}<k$ \red{such that $ \mathfrak{N}(\signalset,\epsilon)\leq \epsilon^{-k}$ for all $\epsilon<\min(\review{\frac{96}{15}} \frac{k}{m \sqrt n} \log(\frac{\review{5} m \sqrt n}{ \review{32} k}),\epsilon_0)$}. \review{For $m>\frac{18 k}{\sqrt{n}}$}, the cardinality of the measurement set is bounded as 
% $$
% \textstyle |\sign{A \signalset}|  \leq \left(\frac{m \sqrt n}{\review{6}k}\right)^{\review{6}k}
% $$ 
% with probability at least $1 - \review{(\frac{32}{5 m \sqrt{n}})^{\frac{6}{5}}}$. % \LJ{could be 1/10 instead of 3/32 ;-)}
% \end{theorem}
\noindent %\LJ{The following text must be adapted.} 
The proof is included in \Cref{app: sample complexity}. This result depends on the square root of the ambient dimension $\sqrt{n}$ due to the application of~\Cref{lemma: lemma laurent} and can be suboptimal for some signal sets. For example, the bound in~\Cref{eq: n uos} avoids this dependency for the case where $\signalset$ is a union of subspaces. %Moreover, the dependency of the probability in $m$  can be increased by loosening the bound on $N$ (see the proof for more details).}
%\LJ{Acknowledge the presence of $n$ due to the Lemma x, as well as the fact that the exponent $-2$ is arbitrary and could be increased.}

\review{In the setting where we observe measurements through $\ntransf$ independent forward operators, we sum the number of intersected cells for each operator, so that with probability exceeding $1-G\xi$ for $0<\xi<1$ (by a union bound), the number of different binary measurement vectors is then bounded by
$$
\ts N =  \cl O\Big(\ntransf \left(\frac{m \sqrt n}{k}\right)^{5 k}\Big)
$$
with a hidden multiplicative constant depending on $\xi$.}
Similarly to \Cref{eq: n uos}, this bound scales exponentially only in the model dimension $k$ but not in the number of measurements $m$ or operators $\ntransf$. In the setting of a single operator and a $k$-dimensional invariant signal set, we have the upper bound \review{$\ts N =  \cl O\Big( \left(\frac{m \sqrt n}{k}\right)^{5 k}\Big)$}.

\section{Learning Algorithms}\label{sec: algorithms}

In this section, we present a novel algorithm for learning the reconstruction function $f:(y,A)\mapsto x$ from $N$ binary measurement vectors $\{(y_i, A_{g_i})\}_{i=1}^{N}$, which is motivated by the analysis in~\Cref{sec: model ident}.
We parameterize the reconstruction function using a deep neural network with parameters $\theta\in\R{p}$.
The learned function can take into account the knowledge about the forward operator by simply applying a linear inverse at the first layer, \ie $f_{\theta}(y,A)=\tilde{f}_{\theta}(A^{\top}y)$, or using more complex unrolled optimization architectures~\citep{monga2021algorithm}. 

In the case where we observe measurements associated with $\ntransf$ different forward operators, we propose the SSBM loss 
\begin{align} \label{eq:multop loss}
    \argmin_{\theta\in \R{p}} &\; \sum_{i=1}^{N} \Big[\, \mathcal{L}_{\text{MC}}\left(y_i,A_{g_i}\hat{x}_{\theta,i}\right)\ +\ \alpha \sum_{s\neq g_i}\| \hat{x}_{\theta,i}-  f_{\theta}( \sign{A_s\hat{x}_{\theta,i}},A_s) \|_2^2\,\Big], 
\end{align}
where $\hat{x}_{\theta,i}=f_{\theta}(y_i,A_{g_i})$, 
the cost $\mathcal{L}_{\text{MC}}\left(y_i,A_{g_i}\hat{x}_{\theta,i}\right) \geq 0$ enforces \emph{measurement consistency} (MC), \ie require that $y_i=\sign{A_{g_i}\hat{x}_{\theta,i}}$, and $\alpha\in \mathbb{R}_{+}$ is a hyperparameter controlling the trade-off between the two terms involved.  
In the setting where we have a single operator \review{and the set $\signalset$ is invariant to a group of transformations $\{T_g\}_{g=1}^{\ntransf}$ such as rotations or translations}, we aim to learn a reconstruction function $f_{\theta}:y\mapsto x$ (we remove the dependence of $f_{\theta}$ on $A$ to simplify the notation) via the following self-supervised loss:
\begin{align} \label{eq:ei loss}
    \argmin_{\theta\in \R{p}} &\; \sum_{i=1}^{N} \Big[\, \mathcal{L}_{\text{MC}}\left(y_i,A\hat{x}_{\theta,i}\right)  +\alpha \sum_{g=1}^{\ntransf}\| T_g\hat{x}_{\theta,i}-  f_{\theta}( \sign{A T_g\hat{x}_{\theta,i}}) \|_2^2\,\Big], 
\end{align}
where $\hat{x}_{\theta,i}=f_{\theta}(y_i)$  and $\alpha\in \mathbb{R}_{+}$. %The first term enforces measurement consistency, whereas the second term enforces equivariance of the imaging system~\citep{chen2021equivariant}, \ie $T_g f_\theta(Ax)=f_\theta(A T_gx)$ for all transformation matrices $g=1,\dots,\ntransf$.
In practice, we minimize~\Cref{eq:multop loss} by mini-batching approaches (\eg stochastic gradient descent) by using \red{sampling one out of the $\ntransf$ operators} at random per batch.
In both cases, we choose the measurement consistency term to be the logistic loss, \ie
\begin{equation}\label{eq:logistic}
\mathcal{L}_{\text{MC}}\left(y,\hat{y}\right)= \log \left(1+\exp(-y \circ \hat{y})\right)
\end{equation}
which enforces sign-consistent predictions which are far from zero, as the logistic function tends asymptotically towards zero as $|\hat{y}|\to \infty$. An empirical analysis in~\Cref{sec: experiments} shows that the logistic loss obtains the best performance across various popular consistency losses.

\paragraph{Analysis of the proposed loss} We focus on the multi-operator loss in~\Cref{eq:multop loss}, although a similar analysis also holds for the equivariant setting. \review{The} first term of the loss enforces measurement consistency, \ie requires $y_i=\sign{A_{g_i}f_{\theta}(y_i,A_{g_i})}$ for every $y_i$ in the dataset. However, in the incomplete setting $m<n$, \corr{the simple pseudo-inverse solution 
\begin{equation} \label{eq: pseudo-inverse}
f(y,A_g) = A_g^{\dagger}y
\end{equation}
with $A_g^{\dagger} = A_g^{\top}(A_gA_g^{\top})^{-1}$,}
is measurement consistent for any number of operators $\ntransf$ and training data $N$. Therefore, the first loss does not prevent learning a function $f_{\theta}(y, A_g)$ which acts independently for each operator (as if there were $\ntransf$ independent learning problems). \corr{The second loss \emph{bootstraps} the current estimates $\hat{x}_{i,\theta}=f_{\theta}(y_i,A_{g_i})$ as new ground truth references, mimicking the supervised loss
\begin{equation}
    \label{eq:empirical-self-supervised-cost}
 \sum_{i=1}^N \sum_{s=1}^{\ntransf} \|\hat{x}_{i,\theta} - f_\theta (\sign{A_s \hat{x}_{i,\theta}}, A_s)\|^2,
\end{equation}
in order to enforce consistency across operators.
Importantly, this additional loss avoids the trivial pseudo-inverse solution in~\Cref{eq: pseudo-inverse}, as 
\begin{equation}
    A_g^{\dagger}y - A_s^{\dagger}\sign{A_s A_g^{\dagger}y} \neq 0
\end{equation}
for $g\neq s$ if the nullspaces of $A_g$ and $A_s$ are different, \eg if the necessary condition in~\review{\Cref{cor: necessary condition mG}} is verified.}

\review{
\paragraph{Model identification perspective}
The learning algorithm constructs a discrete approximation of the signal set using the reconstructed dataset, \ie $\cup_{g=1}^G \tilde{\signalset}_g$ where $\tilde{\signalset}_g = f_{\theta}(\mathcal{Y}_g,A_{g})$ for $g=1,\dots,\ntransf$ and $\mathcal{Y}_g$ is the subset of measurement vectors associated with the $g$th operator. From a model identification perspective, the measurement consistency loss ensures that $\mathcal{Y}_g = A_g\tilde{\signalset}_g$
for all $g=1,\dots,\ntransf$. The second loss ensures consistency across all operators, \ie $\tilde{\signalset}_g = f_{\theta}(A_s\tilde{\signalset}_g,A_s)$
for all pairs $s, g \in \{1,\dots,\ntransf\}$, acting as a proxy for $\tilde{\signalset}_g = \tilde{\signalset}_s$. }

\section{Experiments} \label{sec: experiments}

For all experiments, we use measurement operators with entries sampled from a standard Gaussian distribution and evaluate the performance of the algorithms using by computing the average peak-to-signal ratio (PSNR) on a test set with $N'$ ground-truth signals, that is:
\begin{equation}
\textstyle \frac{1}{N'} \sum_{i=1}^{N'} {\rm PSNR}\Big( x'_i, f_\theta(\sign{A_{g_i} x'_i}, A_{g_i})\Big),
\end{equation}
where the PSNR is computed after normalizing the reconstructed image such that it has the same norm as the reference image, \ie 
\begin{equation}
    {\rm PSNR}(x, \hat{x}) = - 20 \log \|x -\hat{x} \frac{\|x\|}{\|\hat{x}\|} \|.
\end{equation}
%\LJ{I guess we should briefly explain (here or after) how we will assess later the quality of the learned $\theta$. below you use the ``test PSNR" which I guess is the average PSNR over a test set. Therefore, we could explain that, while the reconstruction function $f_\theta$ encodes what we have learned from $\cl X$ given its sampling $\{x_i\}_{i=1}^N \subset \cl X$ and the posed inverse problem, we assess how well $f_\theta$ encodes $\cl X$ from a test set $\{x'_i\}_{i=1}^{N'} \subset \cl X$ and estimate
%$$
%\textstyle \frac{1}{N'} \sum_{i=1}^{N'} {\rm PSNR}\Big( x'_i, f_\theta(\sign{A_{g_i} x'_i}, A_{g_i})\Big),
%$$
%or something similar with any other metric than the PSNR.
%}
We choose $f_{\theta}(y,A)=\tilde{f}_{\theta}(A^{\top}y)$ where $\tilde{f}_{\theta}$ is the U-Net network used in~\citep{chen2021equivariant} with weights $\theta$, and train for 400 epochs with the Adam optimizer with learning rate $10^{-4}$ and standard hyperparameters $\beta_1=0.9$ and $\beta_2=0.99$.

\subsection{MNIST experiments}

%\LJ{In this section we should be really clear on what is $\cl X$ in the case of MNIST. Simply saying that its (effective) boxdim is about 12 is not enough. There is again the question of observing a sampling of $\cl X$ instead of $\cl X$.}

We evaluate the theoretical bounds using the MNIST dataset, which consists of greyscale images with $n=784$ pixels and whose box-counting dimension is approximately $k\approx 12$~\citep{hein2005intrinsic}. We use $6 \times 10^{4}$ images for training and $10^{3}$ for testing. 

\begin{figure}[t]
\centering
\includegraphics[width=.8\textwidth]{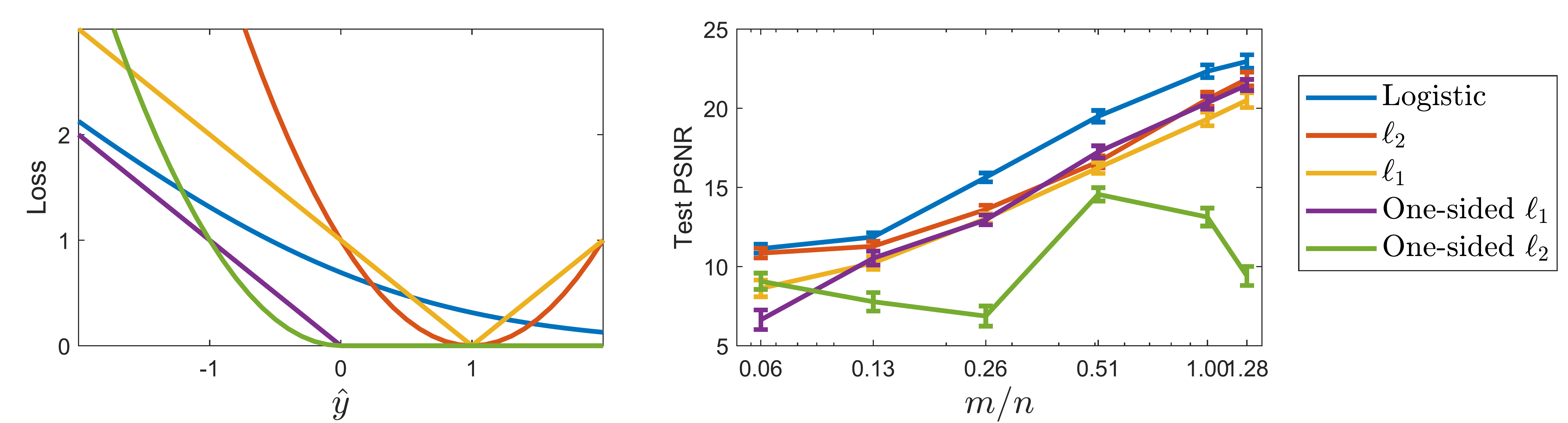}
\caption{Evaluated training losses for enforcing sign measurement consistency $\sign{A\hat{x}}=y$ of reconstructions $f_{\theta}(y)=\hat{x}$. \textbf{Left:} The loss functions are shown for the case $y=1$. \textbf{Right:} Average test PSNR %\LJ{This test PSNR should be explained, see my previous comment.}
of different measurement consistency losses on the MNIST dataset with $\ntransf=10$ operators.}
\label{fig:losses}
\end{figure}

\paragraph{Multiple operators setting.}
We start by comparing the logistic consistency loss in~\Cref{eq:logistic} with the following alternatives:
\begin{itemize}
    %\item 0-1 loss: $\mathcal{L}_{\text{MC}}\left(y,\hat{y}\right)=\| y- \sign{\hat{y}}\|^2$. This loss 
    \item Standard $\ell_p$-loss, $\mathcal{L}_{\text{MC}}\left(y,\hat{y}\right)=\| y- \hat{y} \|_p^p$. As this loss is zero only if $\hat{y}=y$, it promotes sign consistency, $\sign{\hat{y}}=y$ and unit outputs $|\hat{y}|=1$. 
    \item One-sided $\ell_p$-loss, $\mathcal{L}_{\text{MC}}\left(y,\hat{y}\right)=\| \max(-y \circ \hat{y},0) \|_p^p$ where $\circ$ denotes element-wise multiplication and the $\max$ operation is performed element-wise. This loss is zero as long as $\sign{\hat{y}}=y$ regardless of the value of $|\hat{y}|$. 
\end{itemize}
%For each loss we chose the best performance across trade-off parameter $\alpha\in \{0.1,1,10\}$. 
\review{For all losses except the logistic loss, setting the trade-off parameter $\alpha=1$ obtained best results. For the logistic loss, we performed a sweep over different values of $\alpha$ and $m$, finding that the optimal choice of $\alpha$ decreases with $m$ (see \Cref{app:trade-off} for more details). Thus, we set $\alpha=0.1$ for $m<n$ and $\alpha=0.06$ for $m\geq n$.}
\Cref{fig:losses} shows the different losses and the test performance for different values of measurements using $\ntransf=10$ operators. The logistic loss obtains the best performance across all sampling regimes, whereas the one-sided $\ell_2$ loss obtains the worst results.

Secondly, we compare the logistic loss with the following learning schemes:
\begin{itemize}
    \item Linear inverse (no learning), defined as $\hat{x}_i= A_{g_i}^{\top}y_i$. This reconstruction can fail to be measurement consistent~\citep{goyal1998quantized}.
    \item Standard supervised learning loss, defined as $\sum_{i=1}^{N} \| x_i- f_{\theta}(y_i,A_{g_i}) \|^2.$ We also evaluate this loss together with the cross-operator consistency term in~\Cref{eq:multop loss} which we denote as supervised+.
    \item Measurement consistency loss, defined as $\sum_{i=1}^{N} \mathcal{L}_{\text{MC}}\left(y_i,A_{g_i}f_{\theta}(y_i,A_{g_i}) \right)$ using the logistic loss.
    \item 
    The binary iterative hard-thresholding (BIHT) reconstruction algorithm~\citep{jacques2013robust} with a Daubechies4 orthonormal wavelet basis. The step size and sparsity level of the algorithm were chosen via grid search. It is worth noting that the best-performing sparsity level increases as the number of measurements $m$ is increased.
    \item Proposed \proposed~loss in~\Cref{eq:multop loss} using the logistic loss for measurement consistency.
\end{itemize}

\begin{figure}[t]
\centering
 \begin{subfigure}{0.5\textwidth}
     \centering     \includegraphics[width=1\textwidth]{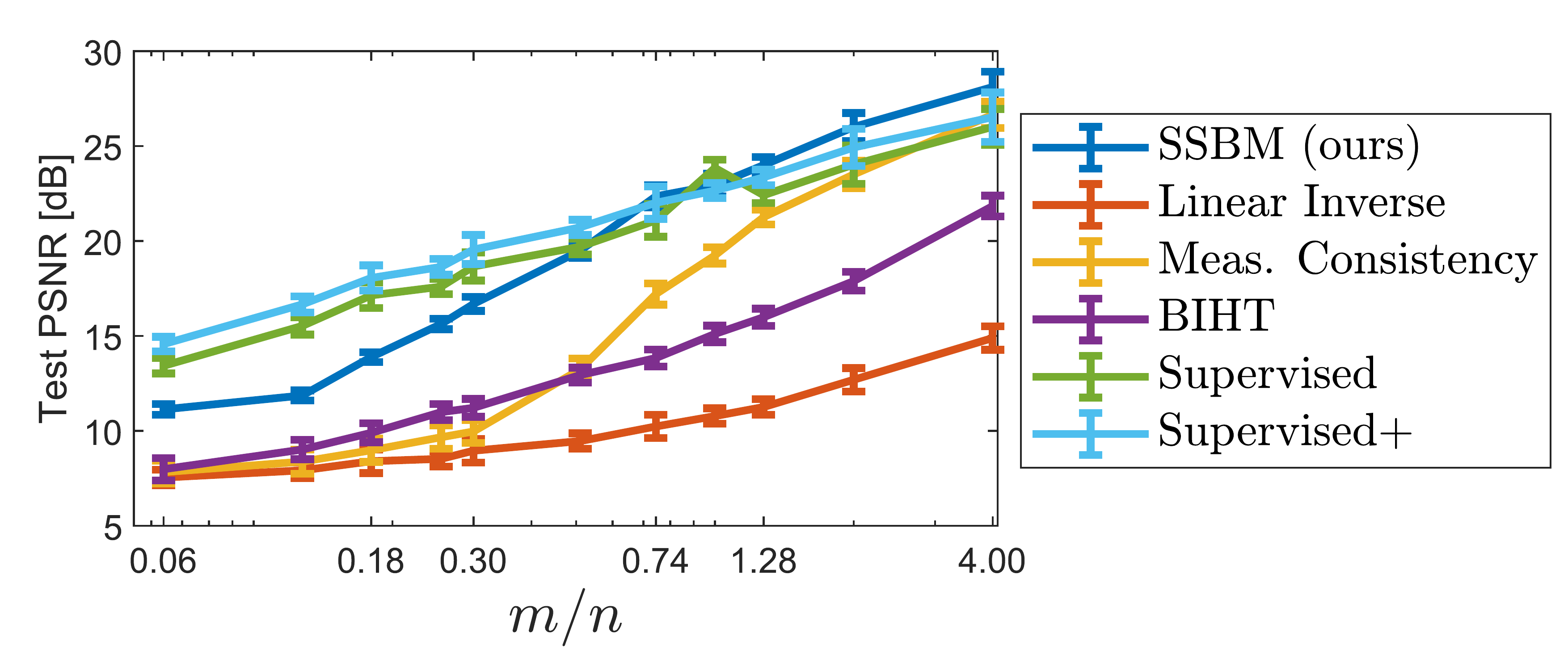}
     %\caption{}
\label{fig: comparison sup} 
 \end{subfigure}
 \begin{subfigure}{0.45\textwidth}
\centering\includegraphics[width=1\textwidth]{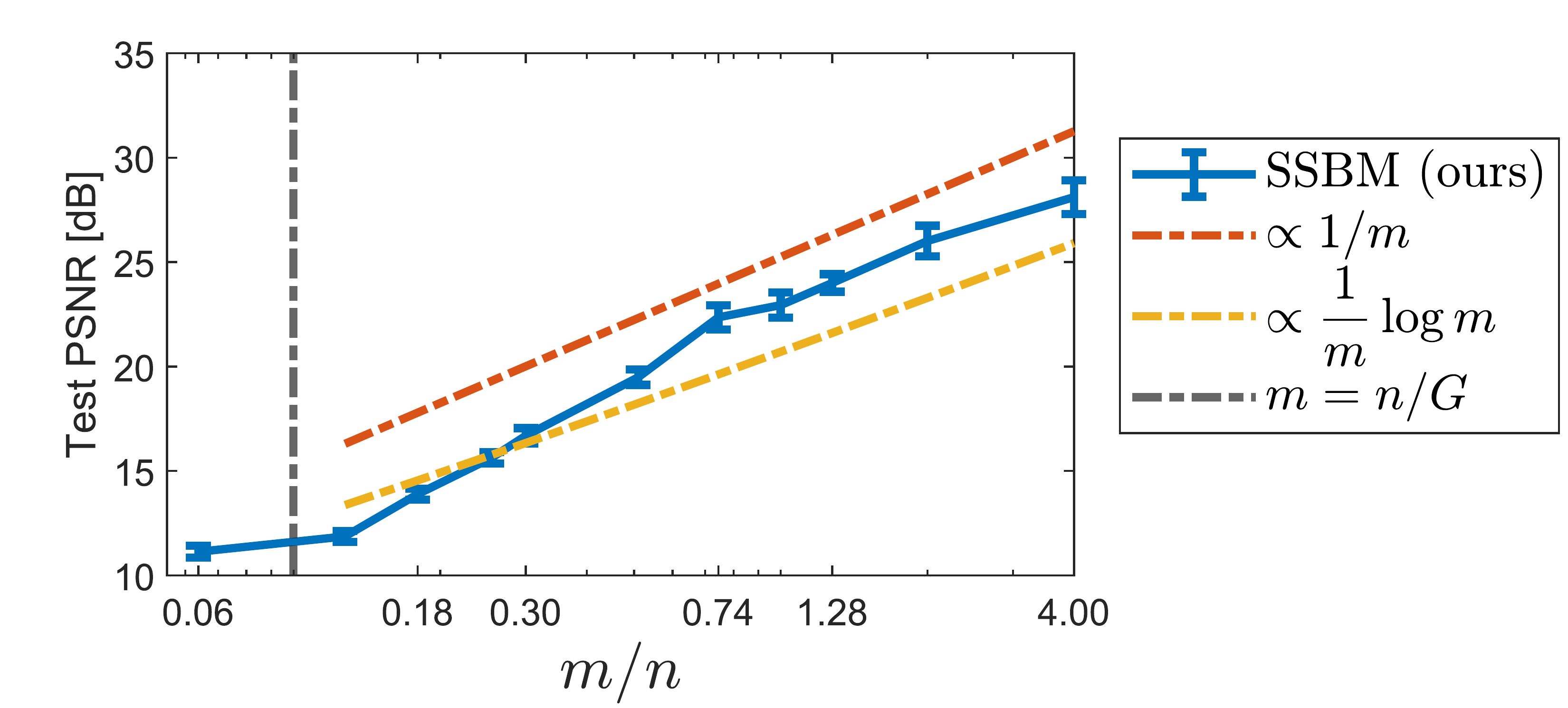}
     %\caption{}
 \end{subfigure}
\caption{\textbf{Left:} Average test PSNR of different supervised and unsupervised algorithms on the MNIST dataset with $\ntransf=10$ operators. \textbf{Right:} The performance of the SSBM method follows closely the bounds \corr{in~\Cref{conj: unsup recov}}. }
\label{fig: allmG10} 
\end{figure}

Test PSNR values obtained for the case of $\ntransf=10$ operators are shown in the left subfigure of~\Cref{fig: allmG10}, where the PSNR in dB is plotted against $m/n$ in log-scale representation. The measurement consistency approach obtains performance similar to simply applying a linear inverse for the incomplete $m/n<1$ setting, whereas it obtains a significant improvement over the linear inverse in the overcomplete case $m/n\geq 1$. This gap can be attributed to the lack of measurement consistency of the linear reconstruction algorithm~\citep{goyal1998quantized}. The proposed loss obtains a performance that is several dBs above the linear inverse and BIHT for all sampling regimes. BIHT relies on the wavelet sparsity prior, which does not capture well enough the MNIST digits.
\proposed~performs similarly to supervised learning as the sampling ratio tends to 1, and perhaps surprisingly, it obtains slightly better performance than supervised learning for $m/n=1.28$. However, adding the cross-operator consistency loss to the supervised method (\ie the method supervised+ in~\Cref{fig: comparison sup}) performs better for all sampling regimes than \proposed. 

The right plot in \Cref{fig: allmG10} compares the performance of the \proposed~with the bounds in~\review{\Cref{prop: reconstruction unsup f}} and~\Cref{conj: unsup recov}. These bounds behave almost linearly in this log-log plot of both the error---through the PSNR---and the log-scale representation of $m/n$. We thus observe a good agreement between the \corr{predictions in~\Cref{conj: unsup recov}} and the performance in practice.

%\footnote{The supervised+ method necessarily performs better than the proposed self-supervised method since it access strictly more information during training.}.

\begin{figure}[t]
\centering
 \begin{subfigure}{0.45\textwidth}
     \centering     \includegraphics[width=.7\textwidth]{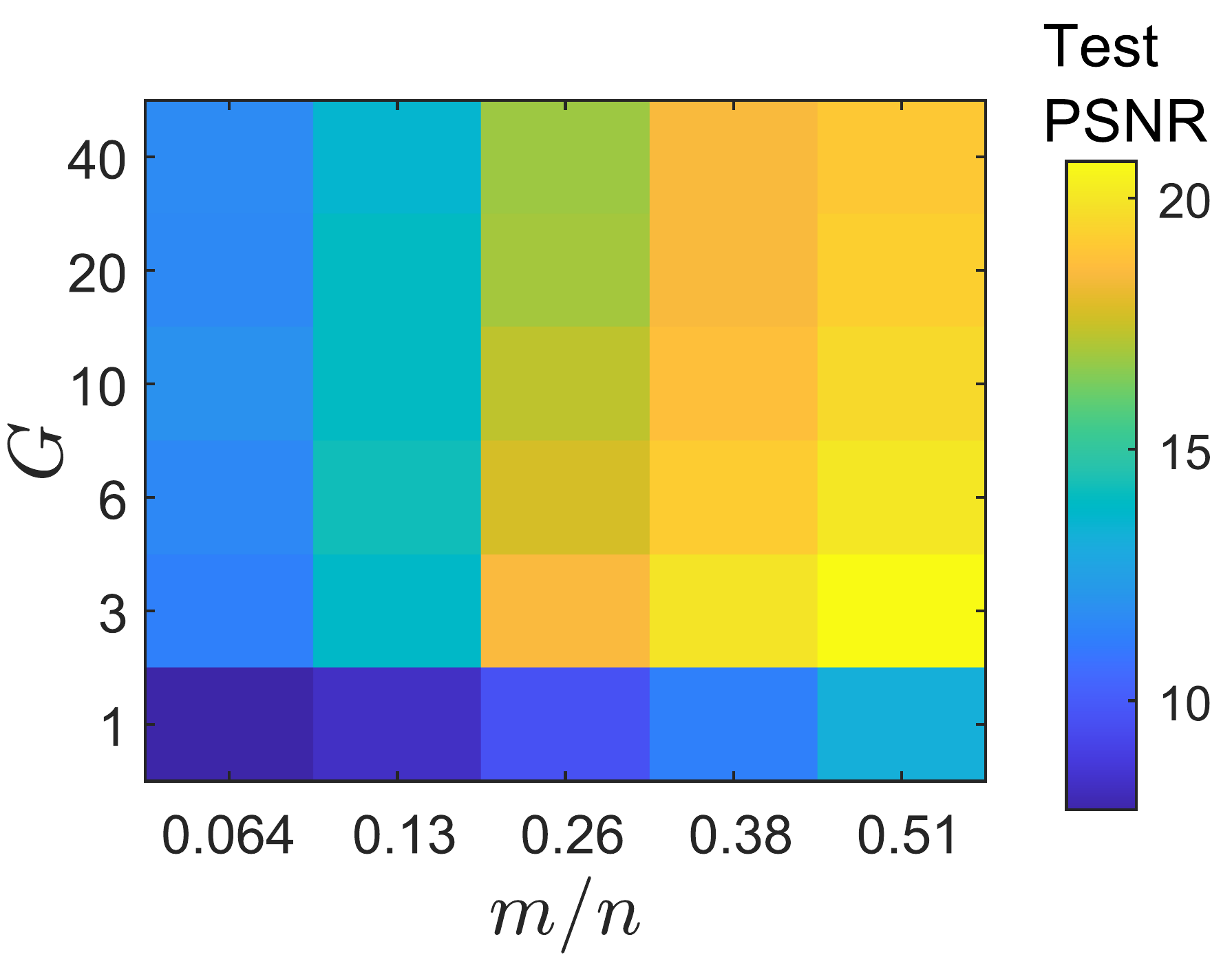}
     \caption{}
 \end{subfigure}
 \hspace{5mm}
 \begin{subfigure}{0.45\textwidth}
\centering\includegraphics[width=.7\textwidth]{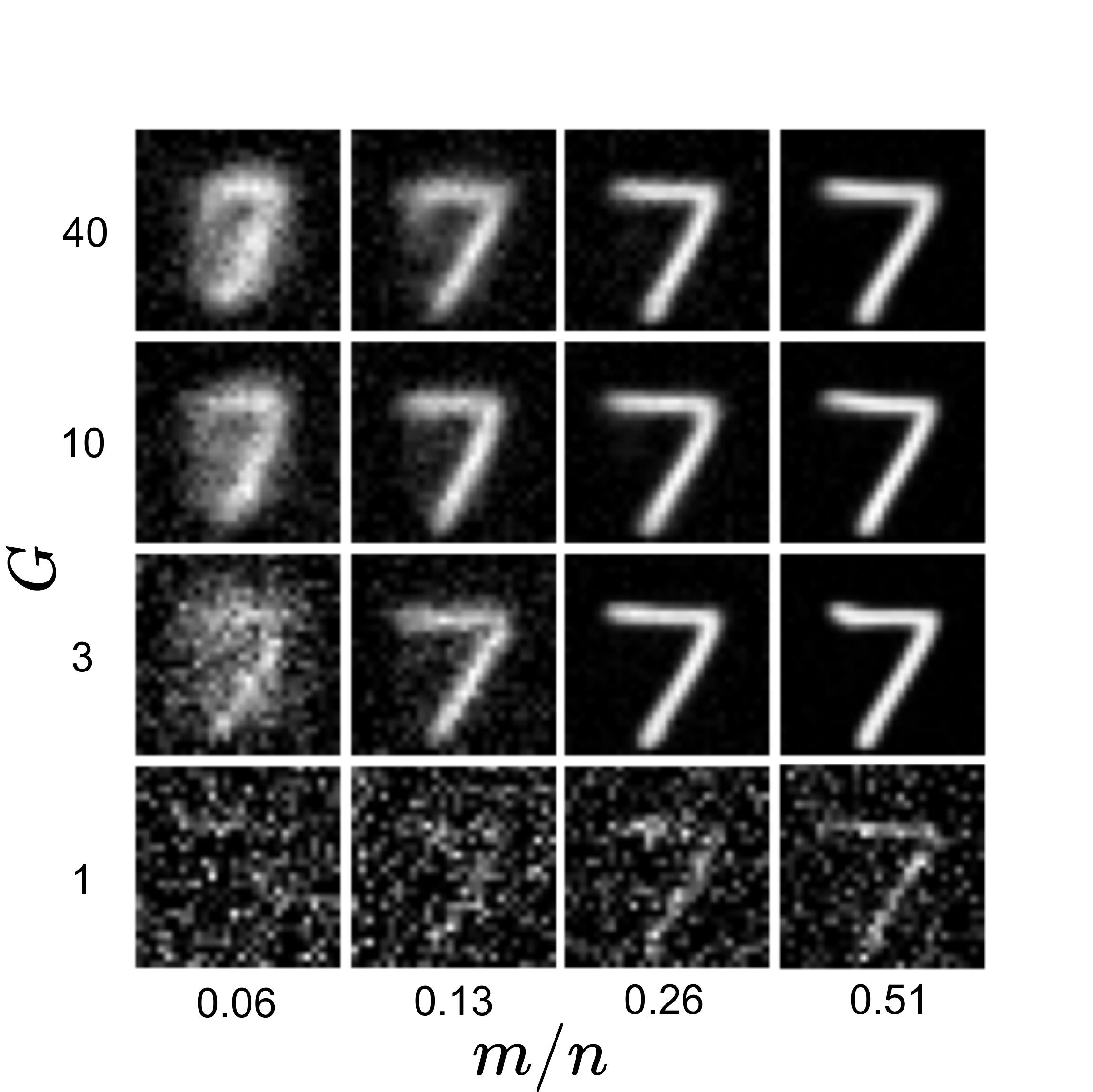}
     \caption{}
 \end{subfigure}
\caption{(\textbf{a}) Average test PSNR and (\textbf{b}) reconstructed test images of the proposed unsupervised method for different numbers of operators $G$ and measurements $m$. }
\label{fig:allmG} 
\end{figure}

\Cref{fig:allmG} shows the average test PSNR and reconstructed images obtained by the proposed self-supervised method for different values of $\ntransf$ and $m$. The method fails to obtain good reconstructions when $\ntransf=1$, as the necessary condition in~\review{\Cref{cor: necessary condition mG}} is not fulfilled.

\begin{figure}[h]
\centering
\includegraphics[width=.8\textwidth]{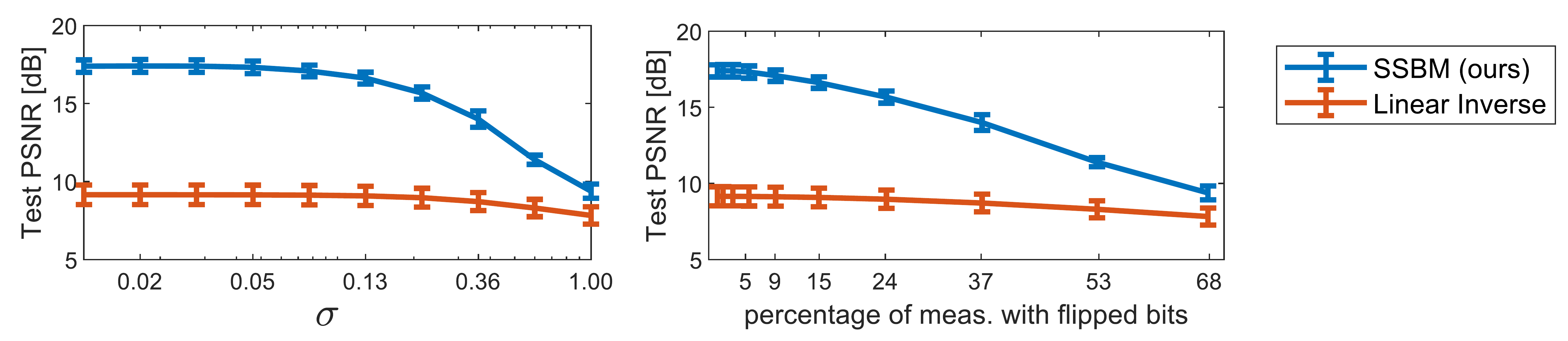}
\caption{\review{Robustness of the proposed learning algorithm to noise in the measurement data. \textbf{Left:} Average test PSNR as a function of the standard deviation of the noise. \textbf{Right:} Average test PSNR as a function of the average percentage of flipped bits per measurement vector.}}
\label{fig: flips} 
\end{figure}

\review{\paragraph{Noisy measurements} In many sensing applications, the sensor data are subject to noise. In the setting of binary measurements, noise affects the measurements by flipping the sign, as the observations can only be $-1$ or $+1$. 
We evaluate the proposed algorithm with $m=274$ measurements and $G=10$ operators and different noise levels, according to the model
\begin{equation}
    y_i = \sign{A_{g_i,i}x_i + \epsilon_i}
\end{equation}
where $\epsilon_i\sim \mathcal{N}(0,I\sigma^2)$ for $i=1,\dots,N$. \Cref{fig: flips} shows the performance of the SSBM algorithm for different values of $\sigma$. The learning algorithm is particularly robust to noise, obtaining a good performance for noise levels up to $\sigma
=0.13$. This noise level translates to having approximately 15\% of bits flipped per measurement vector. It is worth noting that these results indicate that we can expect similarly good performances for other noise distributions (\eg Poisson noise) for a similar average number of bit flips.}

\begin{figure}[t]
\centering
 \begin{subfigure}{0.52\textwidth}
     \centering
     \includegraphics[width=\textwidth]{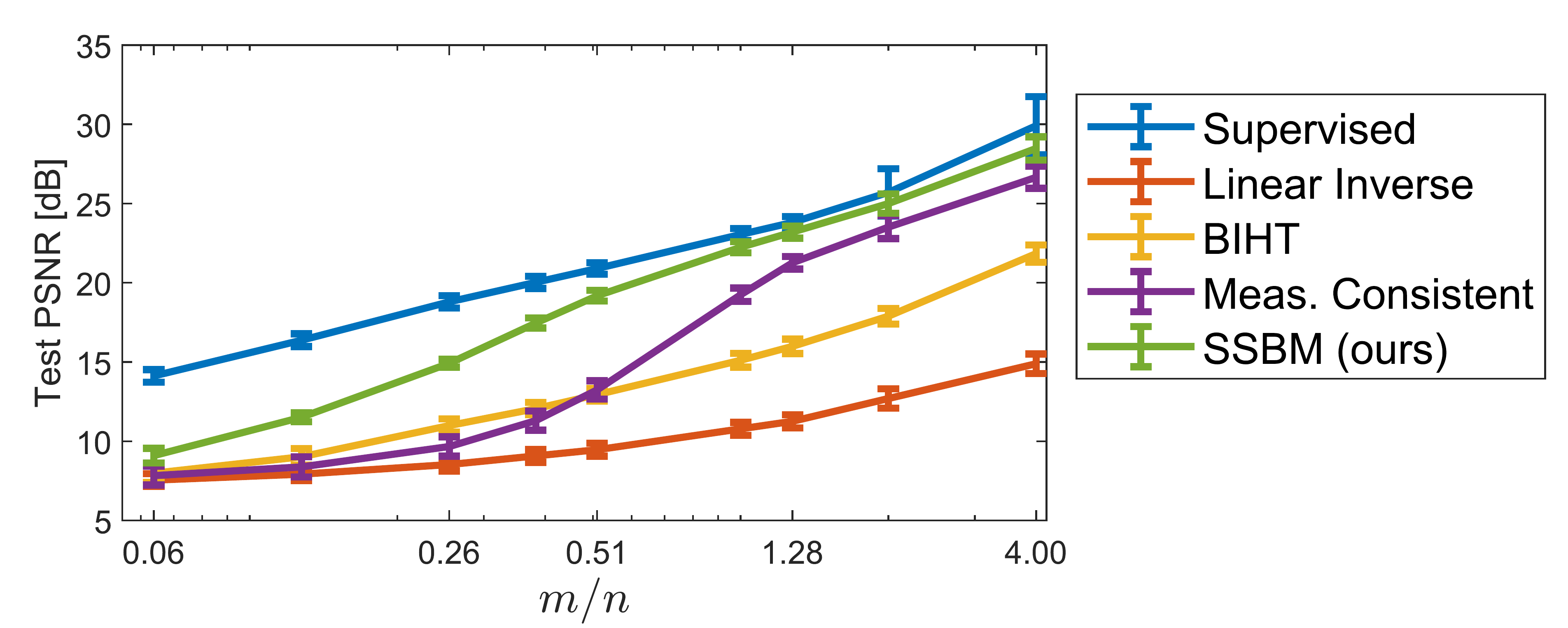}
     \caption{}
 \end{subfigure}
 \hspace{1mm}
 \begin{subfigure}{0.42\textwidth}
     \centering
     \includegraphics[width=1\textwidth]{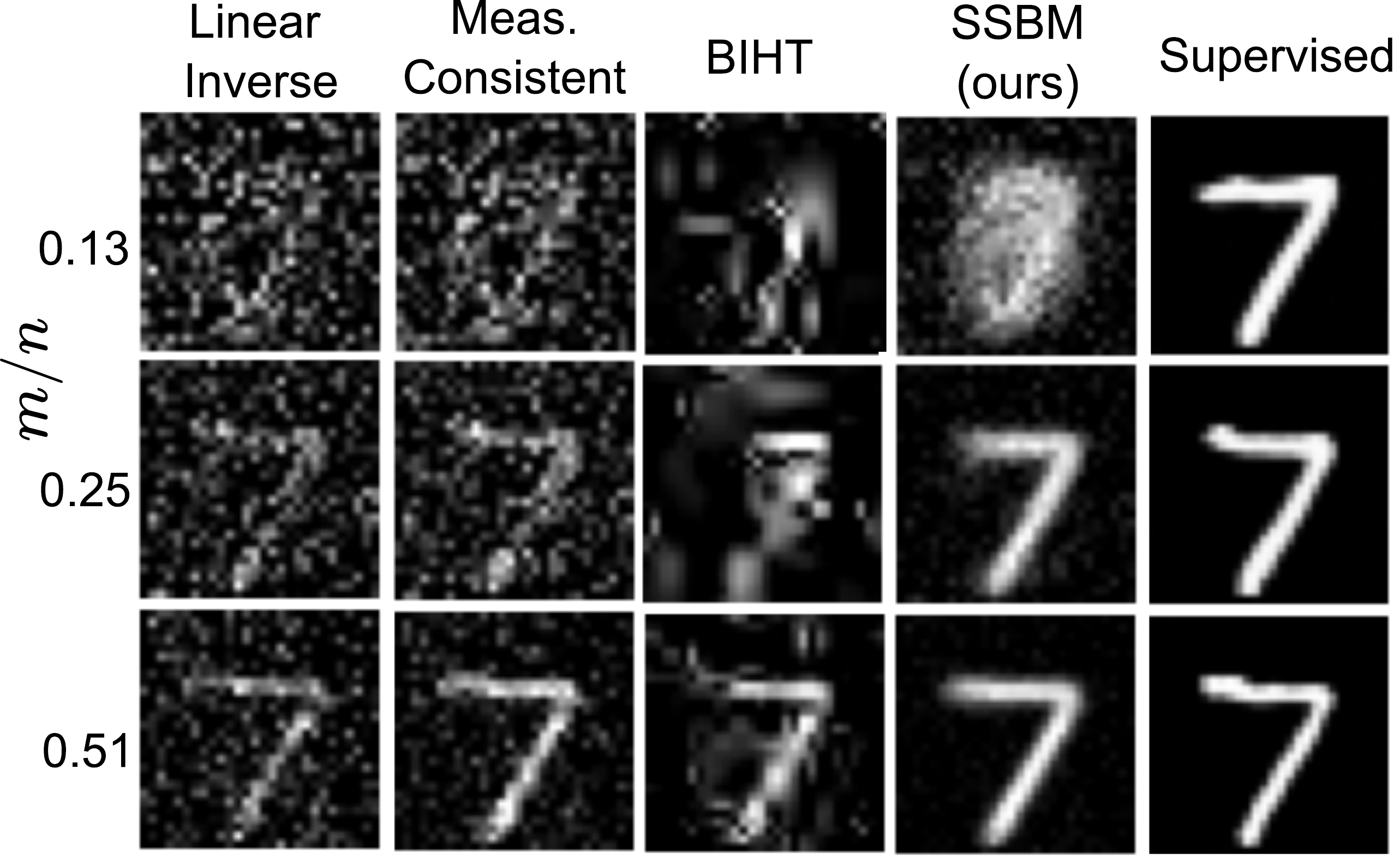}
     \caption{}
 \end{subfigure}
\caption{(\textbf{a})  Average test PSNR and (\textbf{b}) reconstructed test images by the compared algorithms with a single operator $A$ as a function of the undersampling ratio $m/n$. }
\label{fig:allm} 
\end{figure}

\paragraph{Equivariant setting using shifts.}
We evaluate the setting of learning with a single operator by using the unsupervised equivariant objective in~\Cref{eq:ei loss} with 2D shifts as the group of transformations (as the MNIST dataset is approximately shift-invariant). \Cref{fig:allm} shows the average test PSNR %\LJ{to be explained, see above.}
and reconstructed images as a function of the measurements $m$ for various algorithms. The proposed unsupervised method significantly outperforms the linear inverse, BIHT, and the measurement consistent network in all sampling regimes, and performs closely to supervised learning for $m/n>0.4$.

\subsection{Other Datasets}
In order to demonstrate the robustness of the proposed method across datasets, we evaluate the proposed unsupervised approach on the FashionMNIST~\citep{xiao2017online}, CelebA~\citep{liu2015faceattributes} and Flowers~\citep{nilsback2008automated} datasets. The FashionMNIST dataset consists of \review{$6\times 10^{4}$} greyscale images with $28\times 28$ pixels which are divided across $\ntransf=10$ different forward operators. As with MNIST, we use \review{$N=6 \times 10^{3}$} per operator for training and \review{$10^{3}$} per operator for testing. For the CelebA dataset, we use $\ntransf=10$ forward operators and choose a subset of \review{$10^{3}$} images for each operator for training and another subset of the same amount for testing. 
The Flowers dataset consists of 6149 color images for training and 1020 images for testing, all associated with the same forward operator. For both CelebA and Flowers datasets, a center crop of $128\times 128$ pixels of each color image was used for training and testing. 
\Cref{tab: datasets} shows the average test PSNR of the proposed unsupervised method, standard supervised learning, BIHT, and the linear inverse. For BIHT, we use the Daubechies4 orthonormal wavelet basis and optimize the step size and sparsity level via grid search. % sparsity = 20 for fashionMNIST
% Flowers+CelebA sparsity 100 step size 20

\begin{figure}[t]
\centering
\includegraphics[width=.8\textwidth]{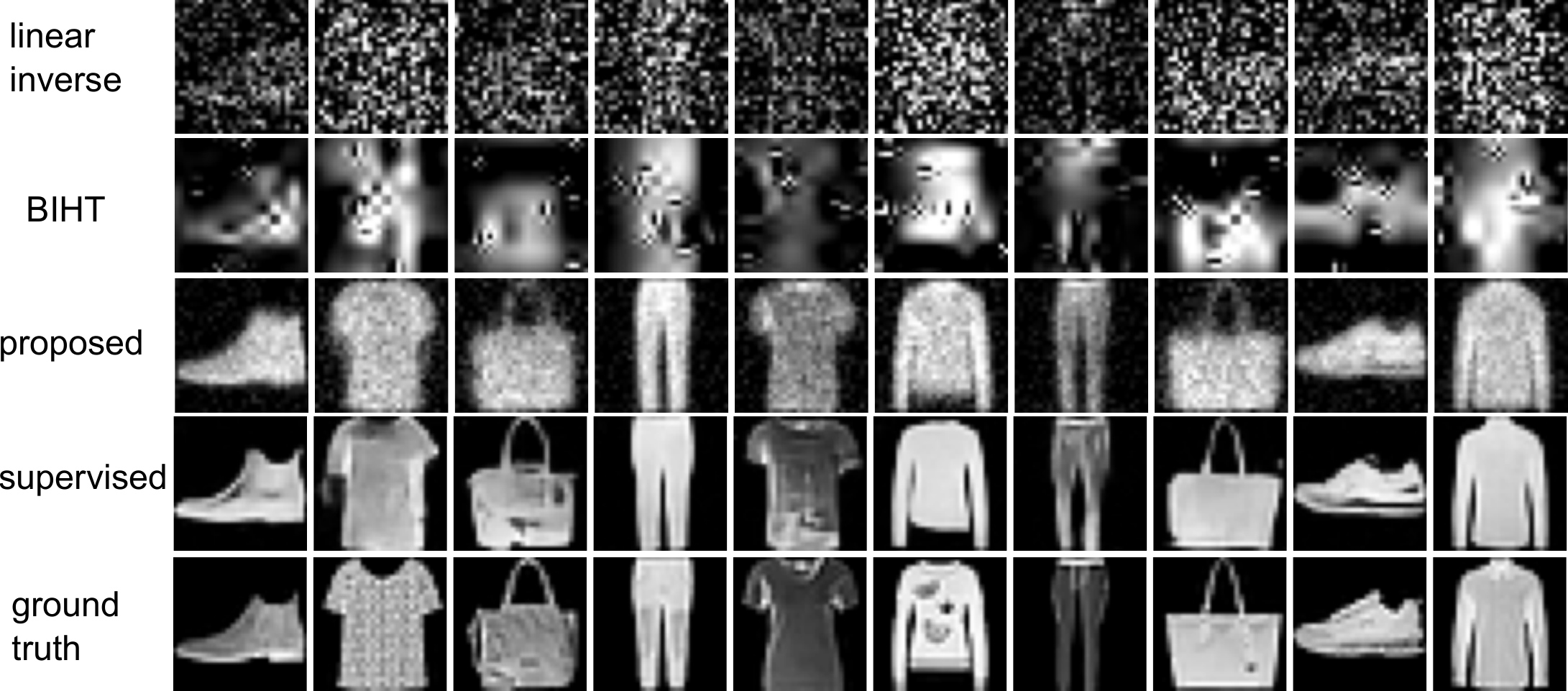}
\caption{Reconstructed test images using the FashionMNIST dataset. Each column corresponds to a \review{test image observed via a}  different forward operator \review{$A_g$}.}
\label{fig:fashion} 
\end{figure}

\begin{figure}[t]
\centering
\includegraphics[width=.9\textwidth]{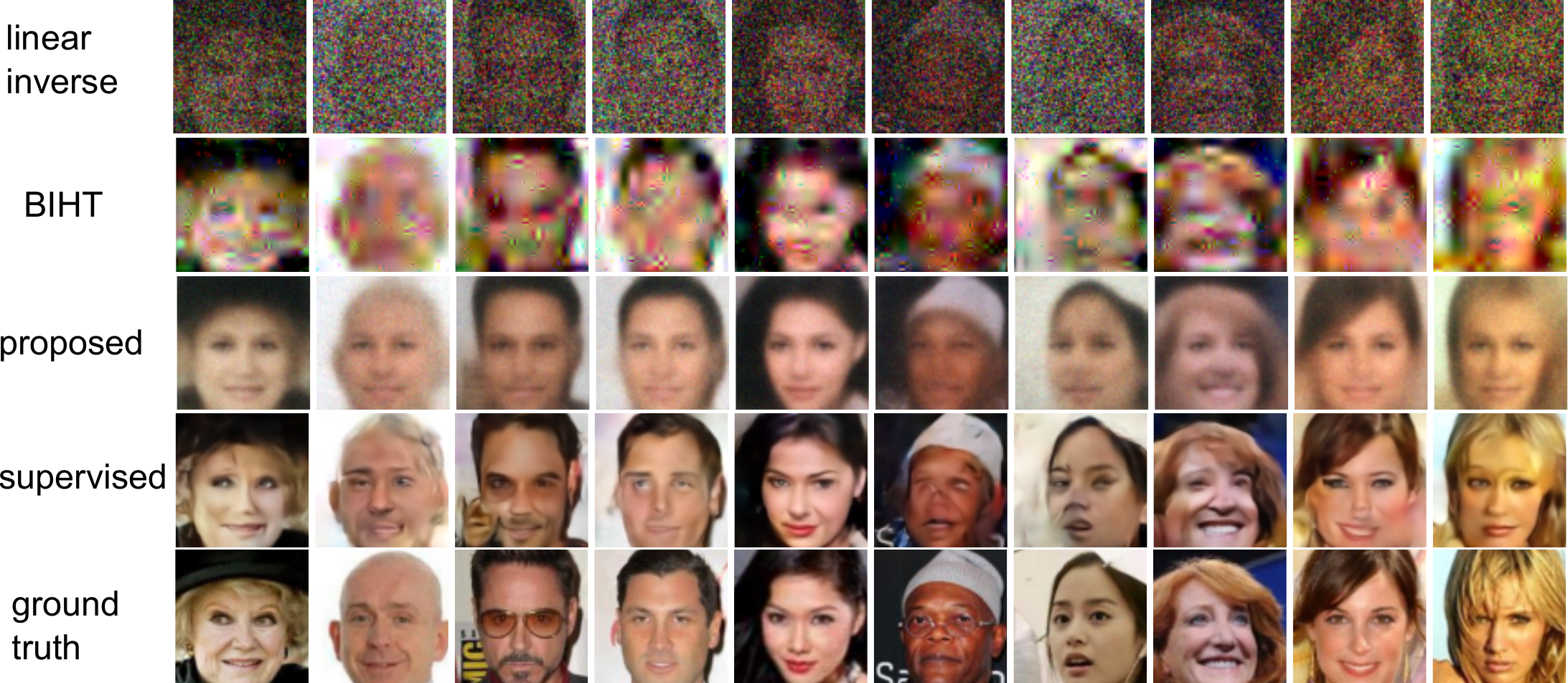}
\caption{CelebA results. Reconstructed test images using the CelebA dataset. Each column corresponds to a \review{test image observed via a} different forward operator \review{$A_g$}.}
\label{fig:celebA} 
\end{figure}

The self-supervised method obtains an average test PSNR which is only 1 to 2 dB below the supervised approach. \Cref{fig:fashion,fig:celebA} show reconstructed test images by the evaluated approaches for each forward operator. The proposed unsupervised method is able to provide good estimates of the images, while only having access to highly incomplete binary information. The supervised method obtains sharper images, however at the cost of hallucinating details, whereas the proposed method obtains blurrier estimates with less hallucinated details.

\begin{table}[t]
\centering \label{tab: datasets}
\scalebox{0.8}{\begin{tabular}{|c|c|c|c|c|c|c|c|}
\hline
Dataset      & $n$   & $m$  & $\ntransf$ & Linear Inverse & BIHT & Supervised       & \proposed (ours)        \\ \hline
FashionMNIST & 784   & 300  & 10        & $6.38\pm 0.23$ & $10.68\pm 0.31$  & $17.63\pm 0.33$  & $16.47\pm 0.22$ \\ \hline
CelebA       & 49152 & 9830 & 10         & $4.81\pm 0.32$& $16.26\pm 0.40$  & $21.59\pm 0.31$ & $19.53\pm 0.3$  \\ \hline
Flowers     & 49152 & 9830 & shifts         & $5.31\pm 0.72$ & $14.62\pm 0.92$  & $18.26 \pm 0.75$ & $16.45\pm 0.71$  \\ \hline
\end{tabular}}
\caption{Average test PSNR in dB obtained by the compared methods for the FashionMNIST, CelebA and Flowers datasets.}
\end{table}

\section{Conclusions and Future Work}

The theoretical analysis in this work characterizes the best approximation of a low-dimensional set that can be obtained from binary measurements. The model identification bounds presented here apply to a large class of signal models, as they only rely on the box-counting dimension, and complement those existing for  signal recovery from binary measurements~\citep{goyal1998quantized,jacques2013robust}. Moreover, the proposed self-supervised loss provides a practical algorithm for learning to reconstruct signals from binary measurements alone, which performs closely to fully supervised learning. This work paves the way for deploying machine learning algorithms in scientific and medical imaging applications with quantized observations, where no ground-truth references are available for training.

We leave the proof of \Cref{conj: unsup recov}, and a study of the effect of noise in the observations and related dithering techniques for future work. Another avenue of future research is the extension of~\Cref{theo: onebit} for the case of operators related through the action of a group.

\section*{Acknowledgments}

Part of this research was supported by the Agence Nationale de la Recherche  (Project UNLIP) and by the Fonds de la Recherche Scientifique – FNRS under Grant T.0136.20 (Project Learn2Sense).

\bibliography{bibliography}

\begin{thebibliography}{34}
\providecommand{\natexlab}[1]{#1}
\providecommand{\url}[1]{\texttt{#1}}
\expandafter\ifx\csname urlstyle\endcsname\relax
  \providecommand{\doi}[1]{doi: #1}\else
  \providecommand{\doi}{doi: \begingroup \urlstyle{rm}\Url}\fi

\bibitem[Alberti et~al.(1991)Alberti, Schirinzi, Franceschetti, and
  Pascazio]{alberti1991time}
G~Alberti, G~Schirinzi, G~Franceschetti, and V~Pascazio.
\newblock Time-domain convolution of one-bit coded radar signals.
\newblock In \emph{IEE Proceedings F (Radar and Signal Processing)}, volume
  138, pp.\  438--444. IET, 1991.

\bibitem[Ball et~al.(1997)]{ball1997elementary}
Keith Ball et~al.
\newblock An elementary introduction to modern convex geometry.
\newblock \emph{Flavors of geometry}, 31\penalty0 (1-58):\penalty0 26, 1997.

\bibitem[Baraniuk et~al.(2017)Baraniuk, Foucart, Needell, Plan, and
  Wootters]{baraniuk2017onebit}
R~Baraniuk, S~Foucart, D~Needell, Y~Plan, and M~Wootters.
\newblock {One-bit compressive sensing of dictionary-sparse signals}.
\newblock \emph{Information and Inference: A Journal of the IMA}, 7\penalty0
  (1):\penalty0 83--104, 08 2017.
\newblock ISSN 2049-8764.
\newblock \doi{10.1093/imaiai/iax009}.
\newblock URL \url{https://doi.org/10.1093/imaiai/iax009}.

\bibitem[Baraniuk \& Wakin(2009)Baraniuk and Wakin]{baraniuk2009random}
Richard~G Baraniuk and Michael~B Wakin.
\newblock Random projections of smooth manifolds.
\newblock \emph{Foundations of computational mathematics}, 9\penalty0
  (1):\penalty0 51--77, 2009.

\bibitem[{Blumensath} \& {Davies}(2009){Blumensath} and
  {Davies}]{blumensath2009uos}
T.~{Blumensath} and M.~E. {Davies}.
\newblock Sampling theorems for signals from the union of finite-dimensional
  linear subspaces.
\newblock \emph{IEEE Transactions on Information Theory}, 55\penalty0
  (4):\penalty0 1872--1882, 2009.
\newblock \doi{10.1109/TIT.2009.2013003}.

\bibitem[Boufounos et~al.(2015)Boufounos, Jacques, Krahmer, and
  Saab]{boufounos2015quantization}
Petros~T Boufounos, Laurent Jacques, Felix Krahmer, and Rayan Saab.
\newblock Quantization and compressive sensing.
\newblock In \emph{Compressed Sensing and its Applications: MATHEON Workshop
  2013}, pp.\  193--237. Springer, 2015.

\bibitem[Bourrier et~al.(2014)Bourrier, Davies, Peleg, P{\'e}rez, and
  Gribonval]{bourrier2014fundamental}
Anthony Bourrier, Mike Davies, Tomer Peleg, Patrick P{\'e}rez, and R{\'e}mi
  Gribonval.
\newblock Fundamental performance limits for ideal decoders in high-dimensional
  linear inverse problems.
\newblock \emph{IEEE Transactions on Information Theory}, 60\penalty0
  (12):\penalty0 7928--7946, 2014.

\bibitem[Chen \& Wu(2015)Chen and Wu]{chen2015amplitude}
Ching-Hsien Chen and Jwo-Yuh Wu.
\newblock Amplitude-aided 1-bit compressive sensing over noisy wireless sensor
  networks.
\newblock \emph{IEEE Wireless Communications Letters}, 4\penalty0 (5):\penalty0
  473--476, 2015.

\bibitem[Chen et~al.(2021)Chen, Tachella, and Davies]{chen2021equivariant}
Dongdong Chen, Juli\'an Tachella, and Mike Davies.
\newblock Equivariant imaging: Learning beyond the range space.
\newblock In \emph{Proceedings of the IEEE/CVF International Conference on
  Computer Vision (ICCV)}, pp.\  4379--4388, October 2021.

\bibitem[Chen et~al.(2022)Chen, Tachella, and Davies]{chen2021robust}
Dongdong Chen, Juli{\'a}n Tachella, and Mike Davies.
\newblock Robust equivariant imaging: a fully unsupervised framework for
  learning to image from noisy and partial measurements.
\newblock In \emph{Proceedings of the IEEE Conference on Computer Vision and
  Pattern Recognition (CVPR)}, 2022.

\bibitem[Davenport et~al.(2014)Davenport, Plan, Van Den~Berg, and
  Wootters]{davenport2014onebit}
Mark~A Davenport, Yaniv Plan, Ewout Van Den~Berg, and Mary Wootters.
\newblock 1-bit matrix completion.
\newblock \emph{Information and Inference: A Journal of the IMA}, 3\penalty0
  (3):\penalty0 189--223, 2014.

\bibitem[Falconer(2004)]{falconer2004fractal}
Kenneth Falconer.
\newblock \emph{Fractal geometry: mathematical foundations and applications}.
\newblock John Wiley \& Sons, 2004.

\bibitem[Goyal et~al.(1998)Goyal, Vetterli, and Thao]{goyal1998quantized}
V.K. Goyal, M.~Vetterli, and N.T. Thao.
\newblock Quantized overcomplete expansions in $\mathbb{R}^{N}$: analysis,
  synthesis, and algorithms.
\newblock \emph{IEEE Transactions on Information Theory}, 44\penalty0
  (1):\penalty0 16--31, 1998.
\newblock \doi{10.1109/18.650985}.

\bibitem[Hein \& Audibert(2005)Hein and Audibert]{hein2005intrinsic}
Matthias Hein and Jean-Yves Audibert.
\newblock Intrinsic dimensionality estimation of submanifolds in
  $\mathbb{R}^d$.
\newblock In \emph{Proceedings of the 22nd international conference on Machine
  learning (ICML)}, pp.\  289--296, 2005.

\bibitem[Jacques et~al.(2013)Jacques, Laska, Boufounos, and
  Baraniuk]{jacques2013robust}
Laurent Jacques, Jason~N. Laska, Petros~T. Boufounos, and Richard~G. Baraniuk.
\newblock Robust 1-bit compressive sensing via binary stable embeddings of
  sparse vectors.
\newblock \emph{IEEE transactions on information theory}, 59\penalty0
  (4):\penalty0 2082--2102, 2013.

\bibitem[Kirmani et~al.(2014)Kirmani, Venkatraman, Shin, Cola{\c{c}}o, Wong,
  Shapiro, and Goyal]{kirmani2014first}
Ahmed Kirmani, Dheera Venkatraman, Dongeek Shin, Andrea Cola{\c{c}}o, Franco~NC
  Wong, Jeffrey~H Shapiro, and Vivek~K Goyal.
\newblock First-photon imaging.
\newblock \emph{Science}, 343\penalty0 (6166):\penalty0 58--61, 2014.

\bibitem[Lehtinen et~al.(2018)Lehtinen, Munkberg, Hasselgren, Laine, Karras,
  Aittala, Aila, et~al.]{lehtinen2018noise2noise}
Jaakko Lehtinen, Jacob Munkberg, Jon Hasselgren, Samuli Laine, Tero Karras,
  Miika Aittala, Timo Aila, et~al.
\newblock Noise2{N}oise.
\newblock In \emph{International Conference on Machine Learning (ICML)}. PMLR,
  2018.

\bibitem[Liu et~al.(2020)Liu, Sun, Eldeniz, Gan, An, and Kamilov]{liu2020rare}
Jiaming Liu, Yu~Sun, Cihat Eldeniz, Weijie Gan, Hongyu An, and Ulugbek~S
  Kamilov.
\newblock {RARE}: Image reconstruction using deep priors learned without
  groundtruth.
\newblock \emph{IEEE Journal of Selected Topics in Signal Processing},
  14\penalty0 (6):\penalty0 1088--1099, 2020.

\bibitem[Liu et~al.(2015)Liu, Luo, Wang, and Tang]{liu2015faceattributes}
Ziwei Liu, Ping Luo, Xiaogang Wang, and Xiaoou Tang.
\newblock Deep learning face attributes in the wild.
\newblock In \emph{Proceedings of International Conference on Computer Vision
  (ICCV)}, December 2015.

\bibitem[Monga et~al.(2021)Monga, Li, and Eldar]{monga2021algorithm}
Vishal Monga, Yuelong Li, and Yonina~C Eldar.
\newblock Algorithm unrolling: Interpretable, efficient deep learning for
  signal and image processing.
\newblock \emph{IEEE Signal Processing Magazine}, 38\penalty0 (2):\penalty0
  18--44, 2021.

\bibitem[Nilsback \& Zisserman(2008)Nilsback and
  Zisserman]{nilsback2008automated}
Maria-Elena Nilsback and Andrew Zisserman.
\newblock Automated flower classification over a large number of classes.
\newblock In \emph{2008 Sixth Indian Conference on Computer Vision, Graphics \&
  Image Processing}, pp.\  722--729. IEEE, 2008.

\bibitem[Oymak \& Recht(2015)Oymak and Recht]{oymak2015near}
Samet Oymak and Ben Recht.
\newblock Near-optimal bounds for binary embeddings of arbitrary sets.
\newblock \emph{arXiv preprint arXiv:1512.04433}, 2015.

\bibitem[Pisier(1999)]{pisier1999volume}
Gilles Pisier.
\newblock \emph{The volume of convex bodies and Banach space geometry},
  volume~94.
\newblock Cambridge University Press, 1999.

\bibitem[Plan \& Vershynin(2013)Plan and Vershynin]{plan2013one}
Yaniv Plan and Roman Vershynin.
\newblock One-bit compressed sensing by linear programming.
\newblock \emph{Communications on pure and Applied Mathematics}, 66\penalty0
  (8):\penalty0 1275--1297, 2013.

\bibitem[Puy et~al.(2017)Puy, Davies, and Gribonval]{puy2017recipes}
Gilles Puy, Mike~E Davies, and R{\'e}mi Gribonval.
\newblock Recipes for stable linear embeddings from hilbert spaces to
  $\mathbb{R}^{m}$.
\newblock \emph{IEEE Transactions on Information Theory}, 63\penalty0
  (4):\penalty0 2171--2187, 2017.

\bibitem[Rencker et~al.(2019)Rencker, Bach, Wang, and
  Plumbley]{rencker2019sparse}
Lucas Rencker, Francis Bach, Wenwu Wang, and Mark~D Plumbley.
\newblock Sparse recovery and dictionary learning from nonlinear compressive
  measurements.
\newblock \emph{IEEE Transactions on Signal Processing}, 67\penalty0
  (21):\penalty0 5659--5670, 2019.

\bibitem[Rudin et~al.(1992)Rudin, Osher, and Fatemi]{rudin1992nonlinear}
Leonid~I Rudin, Stanley Osher, and Emad Fatemi.
\newblock Nonlinear total variation based noise removal algorithms.
\newblock \emph{Physica D: nonlinear phenomena}, 60\penalty0 (1-4):\penalty0
  259--268, 1992.

\bibitem[Tachella et~al.(2022)Tachella, Chen, and
  Davies]{tachella2022unsupervised}
Juli{\'a}n Tachella, Dongdong Chen, and Mike Davies.
\newblock Unsupervised learning from incomplete measurements for inverse
  problems.
\newblock In Alice~H. Oh, Alekh Agarwal, Danielle Belgrave, and Kyunghyun Cho
  (eds.), \emph{Advances in Neural Information Processing Systems}, 2022.
\newblock URL \url{https://openreview.net/forum?id=aV9WSvM6N3}.

\bibitem[Tachella et~al.(2023)Tachella, Chen, and Davies]{tachella2022sensing}
Juli{\'a}n Tachella, Dongdong Chen, and Mike Davies.
\newblock Sensing theorems for learning from incomplete measurements.
\newblock \emph{Journal of Machine Learning Research}, 24\penalty0
  (39):\penalty0 1--45, 2023.

\bibitem[Thao \& Vetterli(1996)Thao and Vetterli]{thao1996lower}
Nguyen~T Thao and Martin Vetterli.
\newblock Lower bound on the mean-squared error in oversampled quantization of
  periodic signals using vector quantization analysis.
\newblock \emph{IEEE Transactions on Information Theory}, 42\penalty0
  (2):\penalty0 469--479, 1996.

\bibitem[Vershynin(2018)]{vershynin2018high}
Roman Vershynin.
\newblock \emph{High-dimensional probability: An introduction with applications
  in data science}, volume~47.
\newblock Cambridge university press, 2018.

\bibitem[Xiao et~al.(2017)Xiao, Rasul, and Vollgraf]{xiao2017online}
Han Xiao, Kashif Rasul, and Roland Vollgraf.
\newblock Fashion-mnist: a novel image dataset for benchmarking machine
  learning algorithms, 2017.

\bibitem[Yaman et~al.(2020)Yaman, Hosseini, Moeller, Ellermann, U{\u{g}}urbil,
  and Ak{\c{c}}akaya]{yaman2020self}
Burhaneddin Yaman, Seyed Amir~Hossein Hosseini, Steen Moeller, Jutta Ellermann,
  K{\^a}mil U{\u{g}}urbil, and Mehmet Ak{\c{c}}akaya.
\newblock Self-supervised learning of physics-guided reconstruction neural
  networks without fully sampled reference data.
\newblock \emph{Magnetic resonance in medicine}, 84\penalty0 (6):\penalty0
  3172--3191, 2020.

\bibitem[Zayyani et~al.(2015)Zayyani, Korki, and
  Marvasti]{zayyani2015dictionary}
Hadi Zayyani, Mehdi Korki, and Farrokh Marvasti.
\newblock Dictionary learning for blind one bit compressed sensing.
\newblock \emph{IEEE Signal Processing Letters}, 23\penalty0 (2):\penalty0
  187--191, 2015.

\end{thebibliography}
\bibliographystyle{tmlr}

\appendix
\section{Technical Lemmas}

%\LJ{Appendix should be split in sections, one per thm/prop proof, and reference to them should be made from the text}
\corr{We begin by introducing some technical results that play an important role in the main theorems of the paper.} We start with a result from~\citep{jacques2013robust}.
\begin{lemma}[Lemma 9 in~\citep{jacques2013robust}]\label{lemma: lemma laurent}
Given $0\leq \epsilon < 1$ and two unit vectors $\tilde{x},\tilde{v} \in \Sp{n-1}\subset \R{n}$ and $a \in \bb R^n$ with $a_{i} \sim_{\iid} \cl N(0,1)$, we have 
\begin{align} \label{eq: disk lemma}
  p_0 &= \Prob \left[ \forall x \in \ball{\tilde{x}}, \forall  v\in \ball{\tilde{v}} \;: \;  \sign{a^{\top} v} = \sign{a^{\top} x}  \right] \geq 1 - d(\tilde{x},\tilde{v}) - \sqrt{n\frac{\pi}{2}}\epsilon \\
p_1 &= \Prob \left[ \forall x \in \ball{\tilde{x}},  \forall v\in \ball{\tilde{v}} \;: \;  \sign{a^{\top} v} \neq \sign{a^{\top} x}  \right] \geq  d(\tilde{x},\tilde{v}) - \sqrt{n\frac{\pi}{2}}\epsilon.
\end{align}
where $d(\cdot,\cdot)$ denotes the angular distance.
\end{lemma}

\review{\textbf{Remark:} The angular distances in \Cref{lemma: lemma laurent} can be translated into Euclidean distances due to the following inequality:
\begin{equation}
   \ts  d(\tilde{x},\tilde{v}) \geq \frac{2}{\pi} \sin \big(\frac{\pi}{2} d(\tilde{x},\tilde{v}) \big) = \frac{1}{\pi}  \| \tilde{x} - \tilde{v} \|.
\end{equation}}

Let $C^0(S)$ denote the set of continuous functions on the set $S$. This lemma has the following corollary:
%\LJ{Useful propositions for Prop.~\ref{thm:max-bits-gauss}}
\begin{corollary}
\label{cor:proba-discontinuous-sign-scp}
Given $\tilde{x} \in \bb S^{n-1}$, $0<\epsilon < 1/2$, $a \in \R{n}$ with $a \sim_{\iid} \cl N(0,1)$, we have 
$$
\ts \bb P\Big[\sign{a^\top \cdot} \notin C^0\big(\ball{\tilde{x}}\cap \bb S^{n-1}\big)\Big] \leq \sqrt{n}\,\epsilon.
$$
\end{corollary}
\begin{proof}
%\LJ{I found this easy proof. Let's see if we keep it or not} 
The proof can be derived from the complement of the event associated with $p_0$ in \Cref{eq: disk lemma} when $\tilde{x}=\tilde{v}$. Here is, however, a simplified proof for completeness. We first observe that $\sign{a^\top \cdot}$ is discontinuous over $\ball{\tilde{x}}\cap \bb S^{n-1}$ iff $|\frac{a^\top \tilde{x}}{\|a\|}| \leq \epsilon$. 
Therefore, by the rotational invariance of the Gaussian distribution \red{we can choose $\tilde{x}=[1,0,\dots,0]^{\top}$} and the probability above amounts to computing 
$$
\textstyle p := \bb P[|\frac{a_1}{\|a\|}| \leq \epsilon] = \bb P[ a_1^2 \leq \epsilon^2 \|a\|^2] = \bb P[  a_1^2 \leq \frac{\epsilon^2}{(1 - \epsilon^2)} (a_2^2 + \ldots + a_n^2)] = \bb E_\xi \bb P[a_1^2 \leq \frac{\epsilon^2}{(1 - \epsilon^2)} \xi],
$$
where $\xi \sim \chi^2(n-1)$. Since $\bb P[a_1^2 \leq \frac{\epsilon^2}{(1 - \epsilon^2)} \xi] \leq \frac{\sqrt 2}{\sqrt{\pi}}  \frac{\epsilon}{\sqrt{1 - \epsilon^2}} \sqrt{\xi}$, and $\bb E_\xi \sqrt \xi \leq \sqrt{\bb E_\xi \xi} \leq \sqrt{n-1} \leq \sqrt{n}$ by Jensen's inequality, we finally get $p \leq \frac{\sqrt 2}{\sqrt \pi} \, \frac{\epsilon}{\sqrt{1 - \epsilon^2}} \sqrt n \leq \frac{2 \sqrt 2}{\sqrt \pi \sqrt 3} \epsilon \sqrt n < \epsilon \sqrt n$.

\end{proof}

\section{Signal Recovery Proof}\label{app: signal recovery}

%We begin with the proof of \Cref{theo: signal recov boxdim}.
\begin{proof}[Proof of \Cref{theo: signal recov boxdim}]
\review{Proving this theorem amounts to showing that the probability of the failure of the event
\begin{equation*}
   \sign{Ax} \review{=} \sign{As} \implies \|x-s\| < \delta 
\end{equation*}
decays exponentially in $m$ provided that 
\begin{equation*}
m \geq  \tfrac{\review{4}}{\delta} \big(2k\log\tfrac{\review{30}\sqrt{n}}{\delta} + \log\tfrac{1}{\xi} \big)
\end{equation*}
holds. In other words, we want to upper bound 
$$
    p_\delta\ :=\ \Prob\left[\exists x_1,x_2 \in \signalset, \|x_1-x_2\|> \delta :\;  \sign{Ax_1} = \sign{Ax_2} \right]
$$
with such an exponential decay. 

As $\bdim{\signalset}<k$, there exist a constant $\epsilon_0\in (0,\frac{1}
{2})$ such that $\mathfrak{N}(\signalset,\epsilon)\leq \epsilon^{-k}$ for all $\epsilon\leq \epsilon_0$. Thus, there is a covering set $Q_{\epsilon}$ of $\epsilon^{-k}$ points, such that for every $x\in\signalset$, there exists a point $q\in Q_{\epsilon}$ which verifies $\|x-q\|<\epsilon$. 

Thanks to this covering, we can upper bound $p_\delta$ as 
$$
\ts p_\delta\ \leq\ \Prob\left[\exists q_1, q_2 \in Q_{\epsilon}, \exists x_1 \in \ball{q_1}, \exists x_2 \in \ball{q_2}, \|x_1-x_2\| > \delta :\;  \sign{Ax_1} = \sign{Ax_2} \right].
$$
However, since $\|x_1-x_2\| > \delta$, we must have $\|q_1-q_2\| \geq \|x_1 - x_2 \| - 2\epsilon > \delta - 2\epsilon$. Therefore, defining $Q_{\epsilon,\delta} = \{(q,q') \in Q_\epsilon \times Q_\epsilon:  \|q-q'\| > \delta - 2\epsilon\}$, the previous upper bound can be enlarged as
$$
\ts p_\delta\ \leq\ \Prob\left[\exists (q_1, q_2) \in Q_{\epsilon, \delta}, \exists x_1 \in \ball{q_1}, \exists x_2 \in \ball{q_2} :\;  \sign{Ax_1} = \sign{Ax_2} \right].
$$

Given a fixed pair $(q_1, q_2) \in Q_{\epsilon, \delta}$,~\Cref{lemma: lemma laurent} shows that for $a\in \R{n}$ drawn from a standard Gaussian distribution
\begin{align}
    \ts \Prob\left[\forall x_1 \in \ball{q_1}, \forall x_2\in \ball{q_2}: \;   \sign{a^{\top} x_1} \neq \sign{a^{\top} x_2} \right] \geq   \frac{1}{\pi}\| q_1 - q_2 \| - \sqrt{\frac{\pi n}{2}}\epsilon > \frac{\delta - 2\epsilon}{\pi} - \sqrt{\frac{\pi n}{2}}\epsilon.
\end{align}
By setting $\epsilon = \epsilon(\delta) = \frac{\delta (4-\pi)}{8+4\pi\sqrt{n\pi/2}}$ and taking the probability of the complementary event  we obtain
\begin{align}
    \Prob\left[\exists x_1 \in \ball{q_1}, \exists x_2\in \ball{q_2} : \;  \sign{a^{\top} x_1} = \sign{a^{\top} x_2}\right] \leq 1-\delta/4.
\end{align}
Therefore, considering the $m$ \iid rows $\{a_i\}_{i=1}^m \subset \bb R^n$ of the matrix $A = (a_1, \ldots, a_m)^\top \in \R{m\times n}$ drawn from a standard Gaussian distribution, we have
\begin{align*}
    &\ts \Prob[\exists x_1 \in \ball{q_1}, \exists x_2\in \ball{q_2} : \;  \sign{A x_1} = \sign{Ax_2}]\\
    &\ts \leq\ \prod_{i=1}^m \Prob[\exists x_1 \in \ball{q_1}, \exists x_2\in \ball{q_2} : \;  \sign{a_i^\top x_1} = \sign{a_i^\top x_2}]\\
    &\ts \leq\ (1-\delta/\review{4})^{m}.
\end{align*}

Applying a union bound to all pairs $(q_1,q_2)\in Q_{\epsilon(\delta),\delta} \subset Q_\epsilon \times Q_\epsilon$, since there are no more that $\binom{|Q_\epsilon|}{2}\leq |Q_{\epsilon}|^2 \leq \epsilon^{-2k}$ such pairs, we obtain
\begin{align}
    \ts p_\delta \leq \Big(\frac{8+4\pi\sqrt{\pi n/2}}{ (4-\pi) \delta}\Big)^{2k}  (1-\delta/4)^{m} 
    \leq \exp \Big( 2k\log(\frac{8+4\pi\sqrt{\pi n/2}}{ (4-\pi) \delta}) - \frac{m \delta}{4} \Big),
\end{align}
where we used $1-\delta/4 \leq \exp(-\delta /4)$ for $\delta> 0$. 

Upper bounding this probability by $0\leq\xi\leq 1$ as in the statement of Thm~\ref{theo: signal recov boxdim} and using the crude bound $(8+4\pi\sqrt{\pi n/2})/(4-\pi) \leq 30 \sqrt{n}$ for $n\geq 1$, we finally obtain $2k\log\frac{30\sqrt{n}}{\delta} + m\frac{\delta}{4} \geq \log\xi$ which gives the sample complexity bound \eqref{eq:thm-consist-cond}
\begin{align}
    \ts m \geq \frac{4}{\delta} \Big(2k\log\frac{30\sqrt{n}}{\delta} + \log\frac{1}{\xi} \Big),
\end{align}
where the condition $\epsilon(\delta) \leq \epsilon_0$ holds if $\delta\leq 30 \epsilon_0 \sqrt{n}$.
}

\end{proof}

\section{Model Identification Proof}\label{app: model ident}

%We continue with the proof of~\Cref{theo: onebit}.
\begin{proof}[Proof of~\Cref{theo: onebit}] We want to \review{identify the condition that $m$, $G$ and $0<\delta<1$ must respect to induce} that $\hat{\signalset}\subseteq\signalset_\delta$ holds with high probability with respect to a random draw of the operators $A_1,\dots, A_{\ntransf}$. Equivalently, we need to show that\review{, for this condition,}
\begin{equation}
    \sign{A_gx_g} = \sign{A_gv}, \quad  \forall g = 1,\dots,\ntransf
\end{equation}
holds for some $v\in \Sp{n-1}\setminus \signalset_\delta $ and some $x_1,\dots,x_{\ntransf}\in \signalset$ with probability at most $\xi$ with respect to a random draw of the Gaussian matrices $A_1,\dots,A_{\ntransf}$.  This proof adapts some of the procedures given in~\citep{jacques2013robust} to our specific setting. We start by bounding this probability for $\epsilon$-balls around vectors $\tilde{v}\in \Sp{n-1}\setminus \signalset_\delta$, $\tilde{x}_1,\dots,\tilde{x}_{\ntransf} \in \signalset$, that is
\begin{align*}
 p_0 &:=\ \Prob  \big[  \exists (x_1, \dots, x_{\ntransf}) \in \ball{\tilde{x}_1} \times \dots \times \ball{\tilde{x}_{\ntransf}}, \exists v\in \ball{\tilde{v}}  \review{\;:\; }  \forall g=1,\dots,\ntransf, \;\sign{A_{g} v} = \sign{A_{g} x_g}\big].
\end{align*}
\review{We first notice that from the independence of the operators $\{A_g\}_{g=1}^G$,
\begin{align*}
 p_0 &\ts \leq\ \prod_{g=1}^{\ntransf} \Prob  \big[  \exists x_g \in \ball{\tilde{x}_g}, \exists v\in \ball{\tilde{v}}  \review{\;:\; }  \sign{A_{g} v} = \sign{A_{g} x_g}\big].
\end{align*}}%
Furthermore, as every row of each operator \review{$A_g$ is \iid as a standard Gaussian random vector $a_g$}, \review{we have}
\begin{align}
p_0\ &\ts \review{\leq}\ \prod_{g=1}^{\ntransf} \Prob  \left[ \exists x_g \in \ball{\tilde{x}_g}, \exists v\in \ball{\tilde{v}} \review{\;:\; }   \sign{a_{g}^{\top} v} = \sign{a_{g}^{\top} x_g} \right]^{m}\\
&\ts = \prod_{g=1}^{\ntransf}\Big(1- \Prob\left[\forall x_g \in \ball{\tilde{x}_g}, \forall v\in \ball{\tilde{v}} \review{\;:\; }  \sign{a_{g}^{\top} v} \neq \sign{a_{g,i}^{\top} x_g} \right]\Big)^{ m} \label{eq: prod}
\end{align}
\review{From~\Cref{lemma: lemma laurent}, we know that 
\begin{equation}
    \ts \Prob\left[\forall x_g \in \ball{\tilde{x}_g}, \forall v\in \ball{\tilde{v}} \review{\;:\; }  \sign{a_{g,i}^{\top} v} \neq \sign{a_{g,i}^{\top} x_g} \right] \geq \frac{1}{\pi}\|\tilde{x}_g - \tilde{v} \| - \sqrt{n\frac{\pi}{2}}\epsilon .
\end{equation}
where the distance $\|\tilde{x}_g - \tilde{v} \|$ can be bounded by $\delta$ to obtain
\begin{equation}
    \ts \Prob\left[\forall x_g \in \ball{\tilde{x}_g}, \forall v\in \ball{\tilde{v}} \review{\;:\; }  \sign{a_{g,i}^{\top} v} \neq \sign{a_{g,i}^{\top} x_g} \right] \geq \frac{\delta}{\pi} - \sqrt{n\frac{\pi}{2}}\epsilon .
\end{equation}}
%where the angular distance can be bounded by the Euclidean distance
%\begin{equation}
%   \ts \pi d(\tilde{x}_g,\tilde{v}) \geq 2 \sin \big(\frac{\pi}{2} d(\tilde{x}_g,\tilde{v}) \big) =   \| \tilde{x}_g - \tilde{v} \|\geq \delta % \| x_g  - v\| - 2\epsilon \geq \| x_g  - v\| - 2\epsilon
%\end{equation}
%for all $x_g \in \ball{\tilde{x}_g}$ and all $v\in\ball{\tilde{v}}$.
\review{\noindent Plugging this into~\Cref{eq: prod} and picking $\frac{\delta}{\pi}-\sqrt{n\frac{\pi}{2}}\epsilon = \frac{\delta}{4}$, which means that 
$$
\ts \epsilon = \epsilon(\delta) = (\frac{4- \pi}{\sqrt{8 \pi^3}})\frac{\delta}{\sqrt n} \leq \frac{1}{18}\frac{\delta}{\sqrt n},
$$
we get 
\begin{align}
\label{eq:large g prod}
\ts p_0 \leq \big(1-\frac{\delta}{\pi}+\sqrt{n\frac{\pi}{2}}\epsilon\big)^{m\ntransf} \leq \exp(-\frac{\delta}{4} mG). 
\end{align}}%
We can extend this result to all vectors $v\in\Sp{n-1}\setminus\signalset_\delta$ and $x_1,\dots,x_{\ntransf}\in \signalset$ by applying a union bound over a covering of the product set $\signalset^{\ntransf}\times (\Sp{n-1}\setminus\signalset_\delta)$. Since we can cover $\signalset$ with $\epsilon^{-k}$ balls with $\epsilon\leq \epsilon_0$ due to the assumption that $\bdim{\signalset}<k$, and also cover $\Sp{n-1}\setminus \signalset_{\delta}$ with 
$(3/\epsilon)^{n}$ balls\review{~\citep{pisier1999volume}}, we have 
\begin{multline}
 \Prob  [  \exists x_1,\dots,x_{\ntransf} \in \signalset, \exists v\in (\Sp{n-1}\setminus \signalset_{\delta}) \review{\;:\; }   \forall g=1,\dots,\ntransf, \;\sign{A_{g} v} = \sign{A_{g} x_g} ] \leq \epsilon^{-k\ntransf}(\epsilon/3)^{-n} p_0 
\end{multline}
\review{Using the bound \eqref{eq:large g prod}, the upper bound on $\epsilon$, and} bounding the resulting probability by $\xi$, \review{we obtain 
\begin{align*}
\xi&\ts \ \geq \epsilon^{-k\ntransf}(\frac{\epsilon}{3})^{-n} \exp(-\frac{\delta}{4} mG) = \exp(k\ntransf \log(\frac{1}{\epsilon}) + n \log(\frac{3}{\epsilon}) -\frac{\delta}{4} mG)\\
&\ts \ \geq \exp \big[ k\ntransf \log(\frac{18 \sqrt n}{\delta}) + n \log(\frac{54 \sqrt n}{\delta}) -\frac{\delta}{4} mG \big].
\end{align*}
Equivalently, $\ts m \geq \frac{4}{\delta}\big[ k \log(\frac{18 \sqrt n}{\delta}) + \frac{n}{\ntransf} \log(\frac{54 \sqrt n}{\delta}) + \frac{1}{\ntransf}\log(1/\xi)\big]$, which holds if   
$$
\ts m \geq \frac{4}{\delta}\big[ (k + \frac{n}{\ntransf}) \log(\frac{54 \sqrt n}{\delta}) + \frac{1}{\ntransf}\log(\frac{1}{\xi})\big].
$$
Recalling, we must have $\epsilon < \epsilon_0$, we observe that this conditions is met if $\delta < 18 \sqrt n\,\epsilon_0$.  
}

% the following inequality
% \begin{align*}
%     &\ts  \\
%     &\ts \epsilon^{-k\ntransf\review{-}n}3^{n} (1-\frac{\delta}{\pi}+\sqrt{n\frac{\pi}{2}}\epsilon)^{m\ntransf} \leq \xi
% \end{align*}
% Solving for $m$ and choosing $\epsilon = \sqrt{\frac{2(4-\pi)^2}{\pi^3 n}}\delta \approx 0.23\sqrt{\frac{1}{n}}\delta$ we get
% \begin{equation}
%     m \geq \frac{1}{\log(1-\frac{\delta}{4})} \left\{ (k+\frac{n}{\ntransf}) \log \frac{5\sqrt{n}}{\delta} + \frac{1}{\ntransf} \log \frac{1}{\xi} + \frac{n}{\ntransf}\log 3   \right\}
% \end{equation}
% %\JT{$\delta< \delta_0 = .23\sqrt{n} \epsilon_0 $}
% for $\delta< 4\sqrt{n} \epsilon_0 $.
% Finally, using the fact that $\log(1-\frac{\delta}{4})\geq \delta/4$, we obtain the desired bound,
% \begin{align}
%     m \geq \frac{4}{\delta} \left\{ (k+\frac{n}{\ntransf}) \log \frac{5\sqrt{n}}{\delta} + \frac{1}{\ntransf} \log \frac{1}{\xi} + \frac{n}{\ntransf}\log 3  \right\}.
% \end{align}
\end{proof}

\paragraph{Derivation of $\delta$.} \review{Here we aim to upper bound the minimum identification error, \ie the minimum value of $\delta$, for a fixed number of measurements $m$.}
The bound in~\Cref{theo: onebit}, that is
\begin{equation}
    % \ts m \geq \frac{4}{\delta} \left\{ (k+\frac{n}{\ntransf}) \log \frac{5\sqrt{n}}{\delta} + \frac{1}{\ntransf} \log \frac{1}{\xi} + \frac{n}{\ntransf}\log 3 \right\} 
    \label{eq:samp-comp-bound}
\review{\ts m \geq \frac{4}{\delta}( (k + \frac{n}{\ntransf}) \log(\frac{54 \sqrt n}{\delta}) + \frac{1}{\ntransf}\log(\frac{1}{\xi}))}
\end{equation}
can be rewritten as a function of $\delta$ as 
\begin{equation} \label{eq: lambert}
    \log (\delta) + \delta a \geq b
\end{equation}
where 
\review{$$
\ts a = \frac{m}{4(k+\frac{n}{\ntransf})},\quad \text{and}\ b =  \log 54\sqrt{n} + \frac{1}{(\ntransf k + n)} \log \frac{1}{\xi}.
$$
Notice that $b \geq 1$, and $a \geq 1$ from \eqref{eq:samp-comp-bound} since $0<\delta < 1$.}
\review{The expression in \Cref{eq: lambert} holds if 
\begin{equation} \label{eq: proxy ineq}
    \ts \delta \geq \frac{1}{a}(\log(a) + b).
\end{equation}
Indeed, \Cref{eq: proxy ineq} implies that $\log(\delta) + \delta a \geq \log(\delta)  +\log (a) + b = \log (a\delta) + b$. However, again from \eqref{eq: proxy ineq}, we get $a\delta \geq \log(a) + b \geq 1$ since $a,b \geq 1$. Therefore, $\log(\delta) + \delta a \geq \log (a\delta) + b \geq b$.}

% Using this notation, the inequality in \Cref{eq: lambert} can be further simplified as
% \begin{align}
%     \log \delta + \delta a &\geq b \\
%     \delta e^{a\delta} \geq e^{b} \\
%     a\delta \exp a\delta \geq a e^b \\
%     a\delta \geq W(a e^b) \\
%     \delta \geq \frac{1}{a} W(a e^b)
% \end{align}
% where the line before the last uses the fact that the inverse of $xe^{x}$ is the Lambert W function denoted as $W(\cdot)$. Since $W(x)\geq \log x - \log \log x$ for all $x\geq e$, we can write 
% \begin{align}
%     \delta &\geq \frac{1}{a} \log(a e^{b}) - \frac{1}{a} (\log \log a + \log b)\\
%     \delta &\geq \frac{1}{a} (\log a+b) 
% \end{align}
%\JT{TODO: verify if removing the loglog term is ok. It seems that Laurent removed it in 
%his analysis.} 
Finally, picking the smallest $\delta$ respecting \eqref{eq: proxy ineq}, we get 
% observing that $a\approx\frac{m}{k+\frac{n}{\ntransf}} $ and $b \approx \log n$
for large $m$, $n$ and $\ntransf$, 
\begin{equation}
\ts \delta = \mathcal{O} \big( \frac{k+\frac{n}{\ntransf}}{m} \log\frac{m n}{k+\frac{n}{\ntransf}}\big)
\end{equation}
which, for $n/\ntransf \ll k$ reads
\begin{equation}
\ts \delta = \mathcal{O} \big( \frac{k}{m} \log\frac{m n}{k}  \big).
\end{equation}

\section{Sample Complexity Proof} \label{app: sample complexity}

\begin{proof}[Proof of \Cref{thm:max-bits-gauss}] %\LJ{A rough idea of the proof machinery is missing here. Hard to tell at first sight why we want to bound the number of discontinuous components.}

 \corr{We aim to bound the number of different cells associated with the binary mapping $\sign{A\cdot}$ which contain at least one element from the signal set $\signalset$, \ie $|\sign{A\signalset}|$. Our strategy consists in obtaining a global bound on the number of discontinuities of the binary mapping \review{(or in other words, sign changes)} over the image of a covering of $\signalset$, which can then be related to the number of different cells that contain at least one element of $\signalset$. }

For $\epsilon < \epsilon_{0}$, let $Q_\epsilon \subset \signalset$ be an optimal $\epsilon$ covering of $\signalset$. If $\bdim{\signalset} < k$, then there exists an $\epsilon_{0}\in (0,\frac{1}{2})$ such that $|Q_\epsilon| \leq \epsilon^{-k}$ for all $\epsilon < \epsilon_{0}$. 
Let us define the number $Z(S)$ of its discontinuous components \review{of the binary mapping $\sign{A\cdot}$} over a set $S \subset \bb S^{n-1}$, \ie 
$$
Z(S) := \big|\{i: \review{\sign{a_i^{\top}\cdot}} \notin C^0(S)\}\big|.
$$

%\LJ{Below, check if we propagate the new constant of \Cref{cor:proba-discontinuous-sign-scp}, paying attention that then $\epsilon \leq \min(1/2, \epsilon_{\signalset})$ for the corollary to hold.}

\review{
From the independence of the $\{a_i\}_{i=1}^m$, we observe that $Z(S) = \sum_{i=1}^m Z_i(S)$ is a binomial random variable, that is the sum of $m$ \iid binary random variables %$Z_i(S) \sim Z_0(S) \in \{0,1\}$
with probability
$$
\ts p := \bb P\Big[\sign{a^\top \cdot} \notin C^0(S)\Big],
$$ %\bb P[Z_0(S) = 1] 
 where $a$ is a standard Gaussian random vector.
From \Cref{cor:proba-discontinuous-sign-scp}, we have  $p \leq \sqrt{n}\epsilon$ for any set of the form $S = S_{q,\epsilon} := \ball{q} \cap \bb S^{n-1}$ with $q \in \bb R^n$. Using Bernstein inequality~\citep{vershynin2018high} on the random variable $Z(S_{q,\epsilon})$ with $\bb E Z(S_{q,\epsilon}) = m p \leq m \sqrt{n}\epsilon$, we obtain
$$
\ts \bb P[Z(S_{q,\epsilon}) > 2 m \sqrt{n}\epsilon] \leq \bb P[Z > 2 \bb E Z] \leq \exp(- \frac{3}{8} m \sqrt{n}\epsilon).
$$
Therefore, since $|Q_\epsilon| \leq \epsilon^{-k}$, we get from a union bound
\begin{equation}\label{eq: prob z cover}
\ts \bb P[\forall q \in Q_\epsilon: \ Z(S_{q,\epsilon}) \leq 2 m \sqrt{n}\epsilon] \leq 1 - \exp(k \log(\frac{1}{\epsilon}) - \frac{3}{8} m \sqrt{n}\epsilon).
\end{equation}

%\hrule

Let us fix $\epsilon$ by setting a failure probability $0 < \xi < 1$ such that $\xi = \exp(k \log(\frac{1}{\epsilon}) - \frac{3}{8} m \sqrt{n}\epsilon)$, \ie 
\begin{equation}
\label{eq:cond-m-xi-epsilon}
\ts  2 m \sqrt{n}\epsilon = \frac{16}{3}\big[\log(\frac{1}{\xi}) + k \log(\frac{1}{\epsilon})\big].
\end{equation} 

This implicitly imposes $\frac{3}{8} m \sqrt{n}\epsilon > k \log(\frac{1}{\epsilon})$, and since $\epsilon < \min(\epsilon_0,1/2)$, we get $\frac{3}{8} m \sqrt{n}\epsilon > k \log 2$, or
\begin{equation}
\label{eq:lbound-epsilon}
\ts \frac{8k\log 2}{3 m \sqrt{n}} < \epsilon.
\end{equation}
Since the left-hand side of \Cref{eq:lbound-epsilon} has to be smaller than $\min(\epsilon_0,1/2)$, we have that $2 k / (m \sqrt{n}) < \min(\epsilon_0,1/2)$  (using the fact that $2 >  \frac{8}{3} \log 2$).}%

\review{For any set $S \subseteq \Sp{n-1}$, the number of cells generated by $\sign{A \cdot}$ in this set cannot exceed 2 to the power of the number of discontinuous components in this mapping, \ie $|\sign{AS}|\leq 2^{Z(S)}$. Thus, with probability $1 - \xi$, and given $q \in Q_\epsilon$, 
$$
\ts |\sign{AS_{q,\epsilon}}| \leq 2^{\frac{16}{3} [\log(\frac{1}{\xi}) + k \log(\frac{1}{\epsilon})]} = (\frac{1}{\xi})^{\frac{16\log 2}{3}} (\frac{1}{\epsilon})^{\frac{16\log 2}{3} k} < (\frac{1}{\xi})^4 (\frac{1}{\epsilon})^{4k}.
$$
Since there are at most $\epsilon^{-k}$ balls in the covering, and using \eqref{eq:lbound-epsilon}, we obtain the bound %\eqref{eq:proba-bound-number-cells-k-dim-space}   
$$
\ts |\sign{A \signalset}| \leq \sum_{q \in \signalset_\epsilon} |\sign{AS_{q,\epsilon}}| < (\frac{1}{\xi})^4 (\frac{1}{\epsilon})^{5k} < (\frac{1}{\xi})^4 \big(\frac{3 m \sqrt{n}}{8k\log 2}\big)^{5k} < (\frac{1}{\xi})^4 \big(\frac{3 m \sqrt{n}}{5k}\big)^{5k}.
$$

We now prove a bound on the expected number of intersected cells. We first observe that 
$$
\ts \bb E |\sign{A \signalset}| \leq \sum_{q \in \signalset_\epsilon} \bb E|\sign{AS_{q,\epsilon}}|,
$$
and, for any set $S \subseteq \Sp{n-1}$, the independence of the random variables $Z_i(S)$ provides
$$
\bb E |\sign{AS}|\leq \bb E\,2^{Z(S)} = \bb E\big[2^{\sum_{i=1}^m Z_i(S)}\big] = \bb E\big[\prod_i 2^{Z_i(S)}\big] = \prod_i \bb E \,2^{Z_i(S)}.   
$$ 
Moreover, considering the previous covering $\cl Q_\epsilon$ of $\cl X$, if $S = S_{q,\epsilon}$, for some $q \in \cl Q_\epsilon$,
$$
\bb E 2^{Z_i(S)} = 2^0 \bb P(Z_i(S) = 0) + 2^1 \bb P(Z_i(S) = 1) = (1-p) + 2 p = 1 + p \leq 1 + \sqrt n \epsilon \leq e^{\sqrt n \epsilon}.  
$$
Therefore, $\bb E |\sign{AS_{q,\epsilon}}|\leq e^{m\sqrt n \epsilon}$, and 
$$
\ts \bb E |\sign{A \signalset}| \leq \epsilon^{-k}e^{m\sqrt n \epsilon} = e^{k \log(\frac{1}{\epsilon}) + m\sqrt n \epsilon}.
$$
The function $k \log(\frac{1}{\epsilon}) + m\sqrt n \epsilon$ is convex in $\epsilon$ and reaches its minimum on $\epsilon = \frac{k}{m\sqrt n}$. Therefore, by setting $\epsilon = k/(m\sqrt n)$ and imposing ${k}/{m\sqrt n} \leq \min(\epsilon_0,1/2)$, we get  
$$
\ts k \log(\frac{m\sqrt n}{k}) + k = k(1+\log(\frac{m\sqrt n}{k})),
$$
which finally gives 
$$
\ts \bb E |\sign{A \signalset}| \leq e^{k(1+\log(\frac{m\sqrt n}{k}))} = (e\frac{m\sqrt n}{k})^k.
$$
}

\end{proof}

%\review{
%\section{Model identification with Noisy Measurements Proof}
%\begin{proof}
%\begin{align}
%&\Prob [ \text{event holds with } m \text{ noisy measurements}]  \\ 
  %& \leq\sum_{m'=1}^{m} \binom{m}{m'} \Prob [\text{at least %} m' \text{ measurements don't flip}] \Prob [ \text{event %holds with } m' \text{ clean measurements}] \\
  %&\leq  \sum_{m'=1}^{m} \binom{m}{m'} \exp(k\ntransf %\log(\frac{18 \sqrt n}{\delta}) + n \log(\frac{54 \sqrt n}{\delta}) -\frac{\delta}{4} m'G)
%\end{align}
%\end{proof}
%}

\begin{figure}[h]
\centering
\includegraphics[width=.4\textwidth]{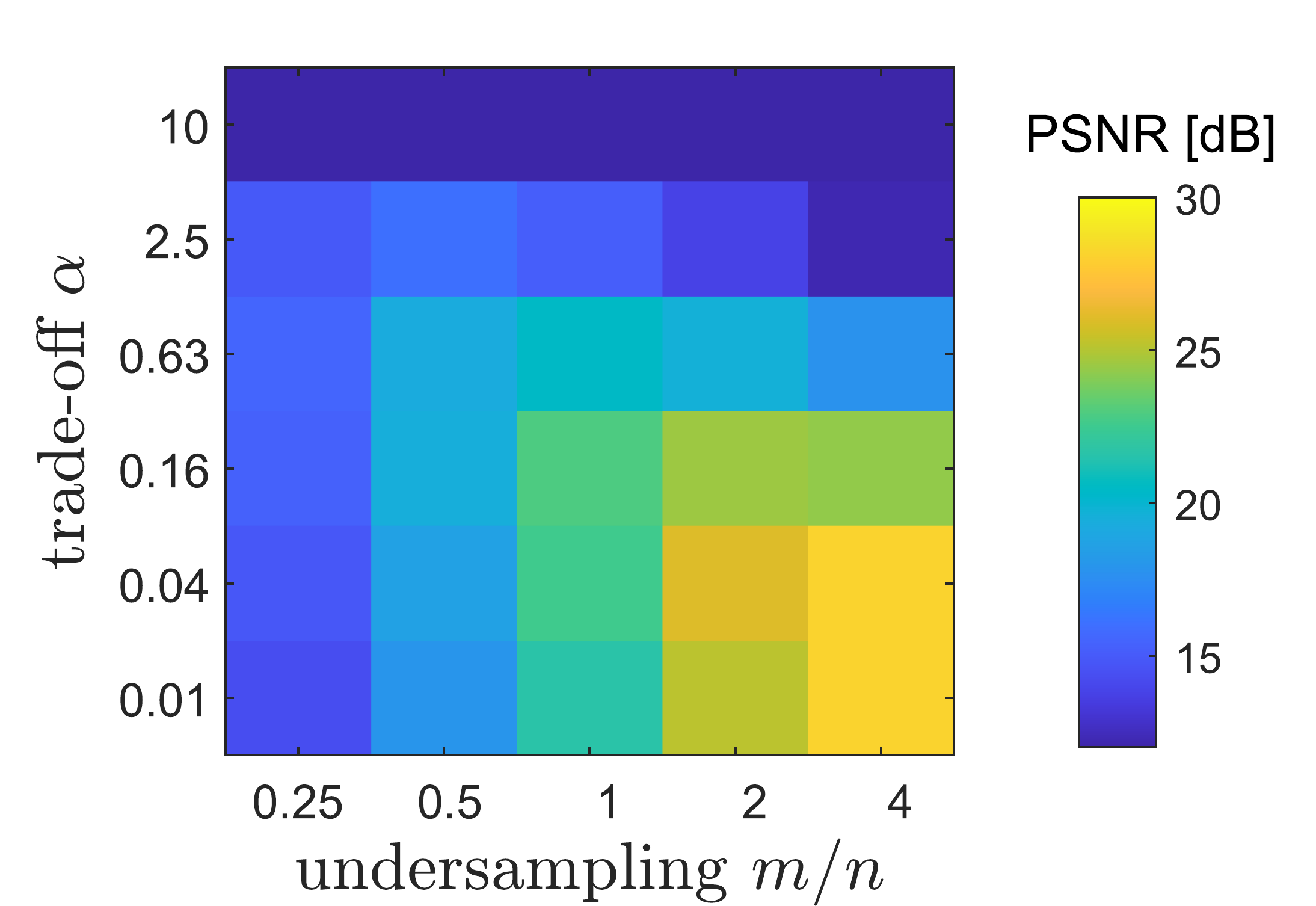}
\caption{\review{Impact of the trade-off parameter $\alpha$ of the SSBM learning algorithm as a function of the sampling ratio $m/n$ network for the MNIST problem with $\ntransf=10$ operators.}}
\label{fig:sweep alpha} 
\end{figure}

\review{
\section{Choice of trade-off parameter}\label{app:trade-off}
We evaluate the impact of the trade-off parameter of the proposed SSBM algorithm for different sampling ratios $m/n$  on the MNIST dataset. In all cases, we use $\ntransf=10$ operators. \Cref{fig:sweep alpha} shows the average test PSNR of the learned networks. The optimal choice of $\alpha$ decreases with the number of measurements $m$. In the experiments, we choose $\alpha=0.1$ if $m<n$ and $\alpha=0.06$ otherwise.}

\end{document}